\documentclass[aps,prd,onecolumn,superscriptaddress]{revtex4}
\usepackage{graphicx}
\usepackage{amsmath}
\usepackage{epsfig}
\usepackage{amsmath,amsfonts,amssymb}
\DeclareGraphicsExtensions{.png,.pdf}
\usepackage{color}

\begin{document}
\baselineskip=12pt
\def\black{\textcolor{black}}
\def\red{\textcolor{black}}
\def\blue{\textcolor{blue}}
\def\green{\textcolor{black}}
\def\be{\begin{equation}}
\def\ee{\end{equation}}
\def\bea{\begin{eqnarray}}
\def\eea{\end{eqnarray}}
\def\orc{\Omega_{r_c}}
\def\om{\Omega_{\text{m}}}
\def\E{{\rm e}}
\def\bearst{\begin{eqnarray*}}
\def\eearst{\end{eqnarray*}}
\def\peleven{\parbox{11cm}}
\def\peffec{\peight{\bearst\eearst}\hfill\peleven}
\def\pspace{\peight{\bearst\eearst}\hfill}
\def\ptwelve{\parbox{12cm}}
\def\peight{\parbox{8mm}}

\title{ISW-Galaxy Cross Correlation:\\ A probe of Dark Energy clustering and distribution of Dark Matter tracers  }

\author{Shahram Khosravi}
\email{khosravi_sh-AT-khu.ac.ir}
\address{Department of Astronomy and High Energy Physics, Faculty of Physics, Kharazmi University, Mofateh Ave., Tehran, Iran}

\author{Amir Mollazadeh}
\email{amirmollazadeh-AT-khu.ac.ir}
\address{Department of Astronomy and High Energy Physics, Faculty of Physics, Kharazmi University, Mofateh Ave., Tehran, Iran}
\vskip 1cm

\author{Shant Baghram}
\email{baghram-AT-sharif.edu}
\address{Department of Physics, Sharif University of
Technology, P.~O.~Box 11155-9161, Tehran, Iran}
\address{School of Astronomy, Institute for Research in
Fundamental Sciences (IPM),
P.~O.~Box 19395-5531,
Tehran, Iran}


\begin{abstract}
Cross correlation of the Integrated Sachs-Wolfe signal (ISW) with the galaxy distribution in late time is a promising tool for constraining the dark energy properties. Here, we study the effect of dark energy clustering  on the ISW-galaxy cross correlation and demonstrate the fact that the bias parameter between the distribution of the galaxies and the underlying dark matter introduces a degeneracy and complications. We argue that as the galaxy's host halo formation time is different from the observation time, we have to consider the evolution of the halo bias parameter.  It will be shown that any deviation from $\Lambda$CDM model will change the evolution of the bias as well. Therefore, it is deduced that the halo bias depends strongly on the sub-sample of galaxies which is chosen for cross correlation and that the joint kernel of ISW effect and the galaxy distribution has a dominant effect on the observed signal. In this work, comparison is made specifically between the clustered dark energy models using two samples of galaxies. The first one is a sub-sample of galaxies from  Sloan Digital Sky Survey, chosen with the r-band magnitude $18 < r < 21$  and the dark matter halo host of mass $M \sim10^{12}M_{\odot}$ and formation redshift of $z_{f}\sim 2.5$. The second one is the sub-sample of Luminous Red galaxies with the dark matter halo hosts of mass $M \sim 10^{13}M_{\odot}$ and formation redshift of $z_{f}\sim 2.0$. Using the evolved bias we improve the $\chi^2$ for the $\Lambda$CDM which reconciles the $\sim$1$\sigma$-2$\sigma$ tension of the ISW-galaxy signal with $\Lambda$CDM prediction. Finally, we study the parameter estimation of a dark energy model with free parameters $w_0$ and $w_a$
in the equation of state $w_{de} = w_0 +w_az/(1+z)$ with the constant bias parameter and also with an evolved bias model with free parameters of galaxy's host halo mass and the halo formation redshift.
\end{abstract}

\maketitle


\section{INTRODUCTION}

For more than a decade now, there has been considerable evidences for the accelerated expansion of the universe. The Hubble-diagram of type Ia Supernova indicates that approximately $70\%$ of the universe is made of an unknown component, dark energy (DE), that accelerates the expansion of universe \cite{Riess:1998cb,Perlmutter:1998np}. The angular power spectrum of Cosmic Microwave Background (CMB) radiation  measured by WMAP \cite{Komatsu:2010fb} and Planck \cite{Ade:2013zuv}, shows that the universe is spatially near flat and DE is needed to explain the CMB temperature fluctuation's power spectrum. On the other hand, the large scale structure (LSS) surveys map the distribution of galaxies in different redshifts and  different angular scales. The statistical properties of galaxy distribution also indicate that the DE plays a major role in the structure formation \cite{Tegmark:2003ud}.
There is an extremely large amount of literature to explain the accelerated expansion of universe at hand. These solutions are mainly categorized in three groups: 1) Cosmological Constant (CC) \cite{Carroll:2000fy}, 2) Modified Gravity (MG) \cite{Nojiri:2006ri,Sotiriou:2008rp, DeFelice:2010aj,Clifton:2011jh}, and 3) Dark energy (DE) models \cite{Peebles:2002gy}.
Accordingly, the cosmological observations that can discriminate between the above solutions become very important \cite{Baghram:2010mc, Mirzatuny:2013nqa, Baghram:2014qja}. Detection of Baryon Acoustic Oscillations (BAO) in matter power spectrum \cite{Eisenstein:2005su,Cole:2005sx,Percival:2007yw,Percival:2009xn}, observation of clusters of galaxies \cite{Vikhlinin:2008ym}, cosmic shear \cite{Fu:2007qq}, and the Lyman alpha forest \cite{Seljak:2004xh} are among the important observations for cosmological studies.
Beside these observations, the Integrated Sachs-Wolfe (ISW) \cite{Sachs:1967er} is a new direction in observational cosmology taken for constraining the cosmological models \cite{Huffenberger:2004tv,Ade:2015dva}.\\
In this work, we study the effect of clustering DE as an extension of the smooth DE quintessential models, on ISW signal, and propose this signal is a promising tool for detecting the clustering effect. However, due to the cosmic variance effect, the ISW signal is difficult to detect directly and can be detected more appropriately via its cross correlation with the late time tracers of dark matter distribution. \cite{Crittenden:1995ak,Boughn:2003yz,Corasaniti:2005pq,Ho:2008bz,Ferraro:2014msa}.
We investigate this effect and show that there is a degeneracy between the DE fluid properties (equation of state and speed of sound) and the bias parameter which relates the clustering of galaxies to the underlying distribution of dark matter. This degeneracy between the cosmological parameters, specially the constant DE equation of state and evolved  bias parameter, is discussed in Schaefer et al. \cite{Schaefer:2009yc}. In this work we go beyond the constant equation of state assumption and propose a theoretical scheme to understand the evolution of bias. In addition, the effect of redshift distribution of dark matter tracers (galaxies) on the ISW-galaxy cross correlation is demonstrated.
It is concluded that the bias parameter and a joint kernel function that encapsulates the effect of the galaxy distributions, change the effectiveness of a DE model on ISW-galaxy cross correlation signal in different redshifts. We study the theoretical ISW-galaxy power spectrum using the evolved bias parameter and investigate the compatibility of models with observation data. For this task, we use the galaxy-galaxy auto-correlation function among other probes and we constrain the DE and bias model parameters.\\
It must be remarked that the future galaxy surveys will expand the sample of dark matter tracers with appropriate sample of galaxies and this will help us probe the effect of DE in a specific redshift, which can be used to impose tighter constraints on the DE parameters via ISW-galaxy correlation.  However, complementary observations are needed to break the degeneracy of bias and DE clustering.\\
This work is structured as below:
In Sec.(\ref{Sec:ISW-g}) a review is presented for the theoretical background of the "Integrated Sachs-Wolfe-galaxy" (ISW-g) cross correlation. In Sec.(\ref{Sec:Clustering}) the clustering of DE models and their effect on the ISW-g signal is studied. In Sec.(\ref{Sec:Bias}) we review the physics of dark matter halo bias and investigate the idea of evolving bias model. In Sec.(\ref{Sec:Res}) we present our results on the ISW-g cross correlation with two samples of galaxies (Luminous Red galaxies (LRG) and Sloan Digital Sky survey(SDSS) magnitude chosen galaxies) for different DE models and different biases. (In first subsection we study the specific models in parameter space and in the second subsection we constrain the parameter space of DE and dark matter halo bias jointly with the addition of galaxy-galaxy auto-correlation data). Finally, Sec.(\ref{Sec:Conc}) is devoted to the conclusion and future remarks. In this paper, we set $\Omega_m h^2=0.132$, $H_0=70.4$ and $n_s=0.95$ for flat $\Lambda$CDM model.

\section{The ISW- galaxy cross correlation}
\label{Sec:ISW-g}
In this section we review the theoretical background for the ISW-galaxy cross correlation signal and its relation to the dark matter power spectrum.
The ISW- galaxy cross correlation idea is used with different classes of galaxies. Pietrobon et al. \cite{Pietrobon:2006gh}, used  WMAP 3 years CMB data with the NRAO VLA Sky Survey (NVSS) radio galaxy, and found further evidence for DE. In \cite{Pietrobon:2006gh}, two specific sound speeds $c_s^2=1$ and $c_s=0$ are discussed. Quasars (QSO) catalog of the SDSS (DR6) is used by Xia et al. \cite{Xia:2009dr} to extract the cross correlation signal. The quasar sample, makes it possible to analysis the behavior of DE in higher redshifts as well. In this study we will use a sub-sample of SDSS galaxies and LRG. \\
In order to study the effect of gravitational perturbations on the CMB photons, we use the perturbed FRW-metric in Newtonian gauge
\be
ds^2=-[1+2\Psi({\bf{x}},t)]dt^2+\left[1+2\Phi({\bf{x}},t)\right]a^2(t)dx^idx_i,
\ee
where $\Psi$ and $\Phi$ are scalar metric perturbations. In the framework of General relativity (GR), the assumption of cosmic fluid with no anisotropic will give $\Psi=-\Phi$.
The Integrated Sachs-Wolfe Effect (ISW) is the imprint of the gravitational potential change on the temperature of the photons free streaming from the last scattering surface. Although, in the matter dominated era the gravitational potential is almost constant, when the universe enters the DE dominated phase the potential gradually becomes shallower. Time dependence of the gravitational potential introduces an effect on temperature perturbations of CMB photons as they fall into and come out of the potential wells. This effect is  related to the change in the matter potential as \cite{Amendola2010}
\be
(\frac{\Delta T}{T})_{ISW}(\theta_x,\theta_y)=-\int_{\eta_i}^{\eta_0} d\eta e^{-\tau}\frac{\partial \psi (\vec{\theta})}{\partial \eta} = \sum_{lm}a_{lm}^{T} Y_{lm}(\hat{n}),
\ee
where temperature anisotropy is a function of two-dimensional position $\hat{n}=(\theta_x,\theta_y)$, $\eta$ is the conformal time, $\tau$ is the optical depth, and the integral is taken from some pre-recombination time $\eta_i$ to the present time $\eta_0$. The second equality comes from the spherical harmonic expansion of temperature change in which $a_{lm}^{(T)}$ are the expansion coefficients. Note that the ISW potential $\psi$ is defined as
\be
\psi({\bf{x}},t)=\Phi({\bf{x}},t)-\Psi({\bf{x}},t).
\ee
It must be remarked that the ISW effect and the weak gravitational lensing are prominent observations that probe both scalar perturbed quantities in geometrical section($\Psi$ and $\Phi$), while the dynamical motion of dark matter tracers and Poisson equation are sensitive to one of the potentials each.
Optical depth $\tau$ is the line of sight integral of the number density of free electrons $n_e$ times the Thompson scattering cross section $\sigma_T$:
\be
\tau=\int_{\eta_i}^{\eta_0} n_{e}\sigma_Td\eta,
\ee
where the integral is taken from the  time of the last scattering surface to present time. In this work we will neglect the optical depth.
Now we can Fourier transform the ISW potential and write the integrals in terms of redshift
\be
(\frac{\Delta T}{T})_{ISW}(\hat{n})=\int  \frac{d^3k}{(2\pi)^3}\int_{z_0}^{z_{CMB}}dz\frac{\partial}{\partial z}\left(e^{i\vec{k}.\vec{r}}\psi_k\right),
\ee
where $\psi_k$ is the Fourier transform of ISW potential.
Assuming a cosmic fluid  with no anisotropic term in its energy-momentum tensor, the ISW potential is related to matter density in dark matter dominated era as
\be
\psi=\frac{3H^2_0\Omega^0_m}{k^2}\frac{D(z)}{D(z=0)}(1+z)\delta^{(0)}_m,
\ee
in which $\delta^{(0)}_m$ and $\Omega^0_m$ are the present day matter density perturbation and the present day density parameter, respectively, $D(z)$ is the growth function of dark matter perturbations normalized to its value in $z=0$, and $H_0$ is the present value of Hubble constant.\\
In order to study the properties of DE, two important points leads us to use the cross correlation of ISW and galaxy distribution. 1) The cosmic variance sets limitations on the accuracy of ISW effect as a
cosmic probe. Therefore, it is more appropriate to extract the signal by using the cross correlation technique with late time tracers of dark matter distribution, such as the distribution of galaxies. 2) The ISW effect is obtained from time evolution of gravitational potential which is related to the dark matter distribution by the Poisson equation. Accordingly, we can study the DE model and its effect on ISW signal via the cross correlation of CMB temperature fluctuations with the matter distribution, as first proposed by \cite{Crittenden:1995ak}.
The matter distribution along the line of sight can be projected on two dimensions as
\be
\delta(\theta_x,\theta_y)=\int dz \frac{dN}{dz}\delta_m(z) = \sum_{lm}a_{lm}^{(g)}Y_{lm}(\hat{n}),
\ee
where $\delta_m(z)$ is the three dimensional matter density contrast, and $dN/dz$ is the selection function which encapsulates the distribution of the galaxies observed by a survey. In the second equality, the two dimensional field of density contrast is expanded in spherical harmonics and $a_{lm}^{(g)}$ are the expansion coefficients.\\
Now we can calculate the angular power spectrum for the cross correlation of coefficients of ISW with galaxy distribution
\be \label{eq:cgt}
C^{gT}_{\ell}\equiv\langle a_{\ell m}^{(g)} a_{\ell m}^{(T)} \rangle=\frac{2}{\pi}\int_0^{\infty}k^2dk I^{ISW}_{\ell}(k)I^{g}_{\ell}(k)P_{m}^{(0)}(k),
\ee
where $P_{\delta}^{(0)}(k)$ is the matter power-spectrum in present time. $I_{\ell}^{ISW}$ and $I_{\ell}^{g}$ represent the specific kernel function of the ISW and galaxy distribution, respectively, which relate the  the matter density  power spectrum to the angular power of the cross correlation. These kernels are obtained as below
\begin{eqnarray}
I_{\ell}^{ISW}(k)&=& \frac{3H_0^2\Omega_{m,0}}{k^2} \int dz\frac{d}{dz}( D(z)(1+z))j_{\ell}[k\chi(z)],\\
\label{eq:Ig}
I_{\ell}^{g}(k)&=&\int dz  b(k,z)\frac{dN}{dz} D(z)j_{\ell}[k\chi(z)],
\end{eqnarray}
where $D(z)\equiv\delta(z)/\delta(0)$ is the growth function. $dN/dz$ is normalized such that $\int dz\frac{dN}{dz}=1$. The spherical Bessel function $j_{\ell}$ depends on the comoving distance $\chi$ while the bias parameter $b(k,z)$ in general is a function of scale and redshift and relates the density contrast of the dark matter to the density contrast of the galaxy distribution.
Dynamics of the DE and any deviation from standard $\Lambda$CDM can change the angular power through the growth function. Furthermore, the relation between the gravitational potential and the matter distribution may be modified in non-$\Lambda$CDM models.
A very important point is that, the low signal to noise ratio of ISW-galaxy cross correlation, makes it difficult to constrain the free parameters of the model. An important observation which can help to resolve this problem is the galaxy-galaxy angular auto-correlation $C^{gg}_l$ which is defined as:
\be
C_l^{gg}=\frac{2}{\pi}\int_0^{\infty} k^2 dk P_{m}^{(0)}(k) I_{\ell}^g(k)I_{\ell}^g(k),
\ee
where $I^g_{\ell}$ is the kernel function of galaxies defined in Eq.(\ref{eq:Ig}). The galaxy-galaxy auto-correlation function will help us to constrain the bias parameter in this approach of the ISW-galaxy cross correlation function. In Sec.(\ref{Sec:Res}) we will use galaxy-galaxy data to constrain the bias.
In the next section, we will study the effect of clustered DE on the signal.
On the other hand the bias parameter introduces another uncertainty beside the cosmological model. The precise determination of the galaxy bias may help to determine any deviation from standard $\Lambda$CDM prediction.
In Sec.(\ref{Sec:Bias}) we will discuss the theoretical aspects of the bias parameter.


\section{Dark Energy Clustering and ISW effect}
\label{Sec:Clustering}


\begin{figure}
\centering
\includegraphics[width=0.5\textwidth]{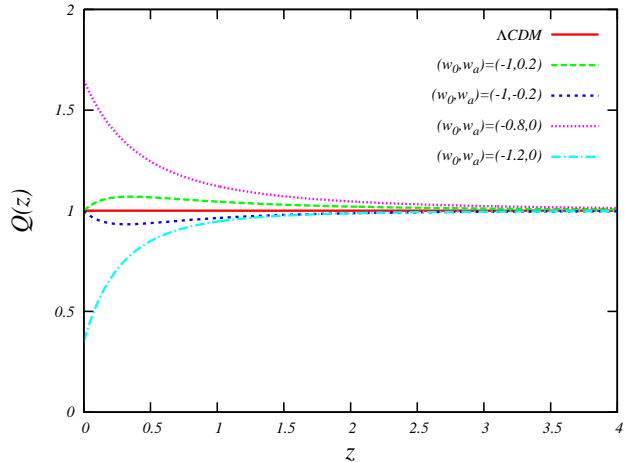}
\caption{The function $Q$ is plotted versus redshift for different models of dark energy. The red solid line indicates $\Lambda$CDM, and the green long-dashed line shows a dark energy model with $(w_0,w_a)=(-1,0.2)$ in CPL parameterizations. The blue dashed line is for a model with $(w_0,w_a)=(-1,-0.2)$. The magenta dotted line  is for the equation of state $(w_0,w_a)=(-0.8,0)$ and cyan dashed-dotted line indicates $(w_0,w_a)=(-1.2,0)$.}
\label{fig:Q}
\end{figure}


In this section we discuss the extension of homogeneous dark energy model to the one with clustering. We will show how the dark energy clustering will change the ISW-g cross correlation through the modified Poisson equation.\\
The dark energy can be considered as a cosmological fluid with a general energy-momentum tensor as below
\be
T_{\mu\nu}=(\rho+P)u_{\mu}u_{\nu}+Pg_{\mu\nu}+q_{\mu}u_{\nu}+q_{\nu}u_{\mu}+\pi_{\mu\nu},
\ee
where $\rho$ and $P$ are the density and pressure of the fluid, $u_{\mu}$ is the four velocity, $q_{\mu}$ is heat transfer vector and $\pi_{\mu\nu}$ is the viscous shear tensor.
First, we assume that the heat transfer and the viscous shear tensor of the DE fluid are zero.
However, we relax the assumption that the DE is a smooth fluid with Jeans length equal to the horizon. Therefore, the fluid can be described by the equation of state $w$ and sound speed $c_s$. In general, the equation of state $w=P/\rho$ can be a redshift-dependent quantity with the cosmological constant ($w=-1$) as a special case. In order to define the sound speed, we consider the fact that the pressure of the fluid can be a function of density and entropy both, ($P=P(\rho,S)$). Thus, we can write
\be
c^2_s=\frac{\delta P(\rho,S)}{\delta \rho}=\frac{\partial P}{\partial \rho}|_S + \frac{\partial P}{\partial S}|_\rho\frac{\partial S}{\partial \rho} = c^2_{s(a)}+c^2_{s(na)},
\ee
where $c^2_{s(a)}$ is the adiabatic sound speed and $c^2_{s(na)}$ is the non-adiabatic sound speed of the fluid. For simplicity we assume  that the extension to the clustering models comes from the adiabatic sound speed term, i.e., $c^2_{s(na)}=0$. In this case, if the sound speed is not equal to unity, it implies that the Jeans length of DE perturbations would be smaller than the horizon. This means that we anticipate DE clusters in sub-horizon scales which will have an effect on the evolution of dark matter  through the field equations.
Now we use the perturbed energy-momentum tensor of the following form \cite{Amendola2010}:
\be
\delta T^\mu_\nu=\rho[\delta(1+c_s^2)u_\nu u^\mu + (1+w)(\delta u_\nu u^\mu + u_\nu \delta u^\mu) +\delta_\nu^\mu c_s^2\delta],
\ee
where $\delta$ is the density contrast. In order to find the evolution of perturbations, we replace the perturbed energy-momentum in Einstein equations $\delta G^{\mu}_{\nu}=8\pi G \delta T^{\mu}_{\nu}$.
Besides the field equations, applying the conservation of energy-momentum tensor,  $T^{\mu}_{\nu;\mu}=0$, to dark matter and DE cosmic fluids, we obtain the continuity and Euler equation in Fourier space:
\begin{eqnarray}
&&\delta'_m+\theta_m + 3\Phi' =0  \\
&&\theta'_m+{\cal{H}}\theta_m-k^2\Psi =0 \\
&&\delta'_{de}+3{\cal{H}}(c_s^2-w)\delta_{de}+(1+w)(\theta_{de}+3\Phi')=0 \\
&&\theta'_{de}+[{\cal{H}}(1-3w)+\frac{w'_{de}}{1+w_{de}}]\theta_{de}-k^2(\frac{c_s^2}{1+w}\delta_{de}+\Psi)=0,
\end{eqnarray}
where $\delta_m$ and $\delta_{de}$ are the dark matter and DE density contrast, $\theta_m$ and $\theta_{de}$ are the Fourier transform of the divergence of each fluid velocity, ${\cal{H}}=aH$ is the conformal Hubble parameter, and the prime is derivative with respect to conformal time. Note that $w$ and $c_s$ are the equation of state and sound speed of DE.
In order to study the distribution of matter in large scales we have to find the time evolution of dark matter density. However, the presence of DE and its properties affects the dynamics and distribution of dark matter through Hubble parameter and the gravitational potential. The gravitational potential is related to the two-fluid perturbed parameters via Poisson equation as below
\be
k^2\Phi=4\pi Ga^2\rho_{m}[\delta_m+\frac{3{\cal{H}}}{k^2}\theta_m+\frac{\rho_{de}}{\rho_m}(\delta_{de}+3(1+w)\frac{{\cal{H}}}{k^2}\theta_{de})].
\ee
In the sub-horizon limit the Poisson equation can be written approximately as
\be \label{eq:pois-mod}
k^2\Phi\simeq 4\pi G\rho_m a^2\delta_m Q,
\ee
where $Q$ is defined as
\be
Q=1+\frac{\rho_{de}\delta_{de}}{\rho_m \delta_m}.
\ee
In the smooth DE models where $\delta_{de}=0$, we get $Q=1$ which is the standard model. Any deviation of parameter $Q$ from unity will be an indication of a deviation from $\Lambda$CDM.
Now, in order to solve for the dynamics of density contrast, we make the assumption that the fluid is barotropic which means that the sound speed is related to equation of state as follows
\be
c_s^2=w-\frac{w'}{3{\cal{H}}(1+w)}.
\ee
In this work we use the Chevallier-Polarski-Linder (CPL) parametrization of the DE \cite{Chevallier:2000qy,Linder:2002et}:
\be \label{eq:CPL}
w_{de}(z)=w_0+w_a(\frac{z}{1+z}),
\ee
where $w_0$ and $w_a$ are free parameters. If $(w_0,w_a)=(-1,0)$, $\Lambda$CDM is recovered.
In Fig.(\ref{fig:Q}), we plot $Q$ versus redshift for different models of DE with CPL equation of state.
The solid redline shows the $\Lambda$CDM, where $Q=1$ and we do not have any deviation from the standard case. In addition, $Q(z)$ is plotted for some different sets of values of the free parameters: $(w_0,w_a)=\{(-1, 0.2), (-1, -0.2), (-0.8, 0), (-1.2, 0)$\}. \\
Considering Eq.(\ref{eq:pois-mod}), it can be seen that $Q$ determines how the Poisson equation is modified. $Q>1$ ($Q<1$) indicates that the same amount of matter perturbation produces more (less) gravitational potential with respect to $\Lambda$CDM. Therefore we anticipate that $Q>1$ ($Q<1$) enhances (diminishes) the amplitude of the ISW-g signal which is related to the matter power spectrum. In the next section, we will show that this is not always the case and in order to study the effect of the DE on the observable quantities we have to know about the tracers of the dark matter distribution as well.\\
In the next step, we have to solve the dynamical equation for the matter density contrast:
\be \label{eq:deltam}
\delta''_m+{\cal{H}}\delta'_m+k^2\Psi+(3{\cal{H}}\Phi'+3\Phi'')=0.
\ee
It is seen that the last term in parentheses can be neglected due to quasi-static approximation. Furthermore, the $k^2\Psi$ term is related to the dark matter and DE density contrast through Eq.(\ref{eq:pois-mod}).
Later, for our theoretical estimation of matter power spectrum we use the solution of Eq.(\ref{eq:deltam}) to extract the growth function of matter distribution in extended DE models and plug it in the matter power spectrum defined as below
\be
P_m(k,z)=A k^{n_s} T^2(k) D^2(z),
\ee
where $D(z)$ is the growth function obtained from solving Eq.(\ref{eq:deltam}). The evolution of potential with scale is introduced in transfer function $T(k)$. In this work we use the Bardeen, Bond, Kaiser, Szalay (BBKS) transfer function \cite{Bardeen:1985tr}. Note that, $A$ and $n_s$ are the amplitude and the spectral index of perturbations.\\
Finally, It must be emphasized that to calculate the ISW-g cross correlation, we need to know the power-spectrum of galaxy distribution. In fact, the specific kernel of ISW-g effect will change in alternative models as below
\begin{eqnarray}
\tilde{I}_{\ell}^{ISW}(k)&=& \frac{3H_0^2\Omega_{m,0}}{k^2} \int dz\frac{d}{dz}( \tilde{D}(z)(1+z) Q(z))j_{\ell}[k\chi(z)]\\
\label{eq:Igtilde}
\tilde{I}_{\ell}^{g}(k)&=&\int dz  b(k,z) \frac{dN}{dz} \tilde{D}(z)j_{\ell}[k\chi(z)],
\end{eqnarray}
where $~\tilde{}~$ indicates the modified kernels due to deviation from $\Lambda$CDM. It is obvious  from Eq.(\ref{eq:Igtilde}), that we must know the relation of galaxy and dark matter  distributions, i.e., the bias parameter, to find the cross-correlation signal. In the next section we will discuss the bias of matter and galaxies and we will show how the DE models change the evolved bias parameter as well. \\ \\


\section{The effect of cosmology on dark matter halo bias}
\label{Sec:Bias}

In this section we will discuss the relation of the galaxy distribution with the underlying dark matter. This is a very important issue in cosmological studies because theoretical models predict the distribution and the growth of dark matter perturbations, while the observations are performed on luminous matter, such as galaxies.
The density contrasts of galaxy and dark matter distributions are related by the bias parameter. The bias parameter in general can be a scale and redshift dependent quantity $b(z,k)$.
The halo bias parameter relates the distributions of dark matter halos to the dark matter itself, while the galaxy bias depends on the Halo Occupation Distribution (HOD) and indicates that in a halo of given mass $M$, how many galaxies with luminosity $L$ exist \cite{Berlind:2001xk}.\\
 Theoretical study and observational investigations are needed to pin down the bias parameter. Although, it is important on its own to study the processes of structure formation, studying the physics of bias is important because it introduces uncertainties in cosmological data interpretation and therefore a probable degeneracy between the bias parameter and the deviation from $\Lambda CDM$ \cite{Mirzatuny:2013nqa}.
Here, we are going to calculate the amount of uncertainty introduced by bias parameter, in which we will mainly focus on the halo bias term and assume the galaxy bias to be approximately one.
The halo bias which can be defined as the number density contrast of halos $n(M,z)$ in the presence of long mode dark matter distribution $\delta_l$, is written as
\be
b(k,z)= \delta_h / \delta_m = (\frac{n(M,z;\delta_l)-\bar{n}(M,z)}{\bar{n}(M,z)} ) / \delta_m ,
\ee
where $\bar{n}(M,z)$ is the number density of structures when long mode perturbation is set to zero. Also, we set the long mode density contrast equal to the matter density contrast in linear regime $\delta_m$. The bias parameter in its simplest definition is just related to the mass of dark matter halo which hosts the observed galaxies. The Press-Schechter bias parameter from the peak background splitting method in Eulerian frame is obtained as \cite{Press:1973iz}
\be
b^{E}_{PS}(M;z)=1+\frac{\nu^2(M,z)-1}{\delta_c},
\ee
where $\delta_c$ is the critical density of spherical collapse and $\nu$ is the height parameter defined as $\nu = \delta_c / \sigma (M)$, in which $\sigma (M)$ is the variance of density perturbation in a window function related to the mass $M$. Of course, the number density of structures is linked to the underlying cosmological model via the matter density variance. In the Press-Schechter formalism the bias parameter is calculated in Lagrangian formalism with an untold assumption that the formation time of a dark matter halo and the observation time are the same. However, that is not quite right and always there is a difference between the formation and observation redshifts.\\
In the case of a possible deviation from $\Lambda$CDM, we must note that the bias parameter evolves differently from formation time to the observation time, in different cosmological models. Therefore, the bias parameter must be considered consistently with the selected model.
 We study the effect of evolution of bias parameter from formation to observation using the continuity and Euler equation for dark and luminous matter \cite{Fry:1992vr,Hui:2007zh, Parfrey:2010uy}.
From Einstein equations in the level of perturbation, we obtain the equations for underlying dark matter:
\begin{eqnarray}
\delta^\prime _m &=&- \nabla . [(1+\delta_m)\textbf{v}_m]\\
\textbf{v}^\prime _m + (\textbf{v}_m . \nabla)\textbf{v}_m + \frac{a^\prime}{a}\textbf{v}_m &=&-\nabla \phi \\
\nabla ^2 \phi &=&4\pi G \bar{\rho}_m a^2 \delta_m,  \label{eq:pois}
\end{eqnarray}
where $\textbf{v}_m$ is the peculiar velocity of matter, $\phi$ is the gravitational potential which can be traced back to metric perturbations, and by $^{\prime}$ we mean the derivative with respect to conformal time. On the other hand, the dynamics of galaxies as point-like particles obey the following equations
\begin{eqnarray} \label{eq:per-g}
\delta^\prime _g &=&- \nabla . [(1+\delta_g)\textbf{v}_g]\\
\textbf{v}^\prime _g + (\textbf{v}_g . \nabla)\textbf{v}_g + \frac{a^\prime}{a}\textbf{v}_g &=&-\nabla \phi,
\end{eqnarray}
where $\delta_g = (n_g - \bar{n}_g)/ \bar{n}_g$ is the galaxy overdensity ($n_g$ is the galaxy number density and $\bar{n}_g$ is its spatial average.), and $\textbf{v}_g$ is the peculiar velocity of the galaxies.
It is seen that the continuity and Euler equations of dark matter and galaxies have the same form with respect to their perturbed quantities while the gravitational potential $\phi$ which is mainly sourced by the dark matter distribution, is exactly the same. So we can write
\begin{equation}
(\textbf{v}_g - \textbf{v}_m)^\prime + \frac{a^\prime}{a}(\textbf{v}_g - \textbf{v}_m)=0,
\end{equation}
which has the solution $(\textbf{v}_g - \textbf{v}_m) \propto \frac{1}{a}$.


\begin{figure}
\centering
\includegraphics[width=0.5\textwidth]{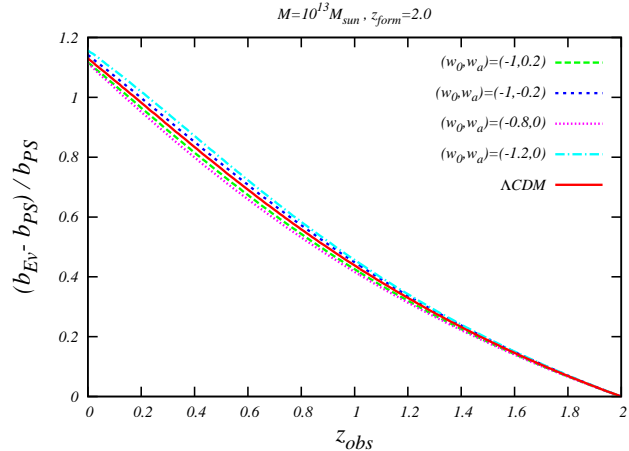}
\caption{The ratio of Evolved bias to Press Schechter bias minus one is plotted versus the observation redshift for different models of dark energy. The red solid line indicates the $\Lambda$CDM, the green long-dashed line shows a dark energy model with $(w_0,w_a)=(-1,0.2)$ in CPL parameterizations. The blue dashed line is for a model with $(w_0,w_a)=(-1,-0.2)$. The magenta dotted line  is for equation of state $(w_0,w_a)=(-0.8,0)$ and cyan dashed-dotted line indicates $(w_0,w_a)=(-1.2,0)$. The dark matter halo is assumed to have a mass of $M =10^{13} M_{\odot}$ as a typical mass of a SDSS LRG catalog galaxies host with formation redshift $z_{f}=2.0$.}
\label{fig:bias-LRG}
\end{figure}


\begin{figure}
\centering
\includegraphics[width=0.5\textwidth]{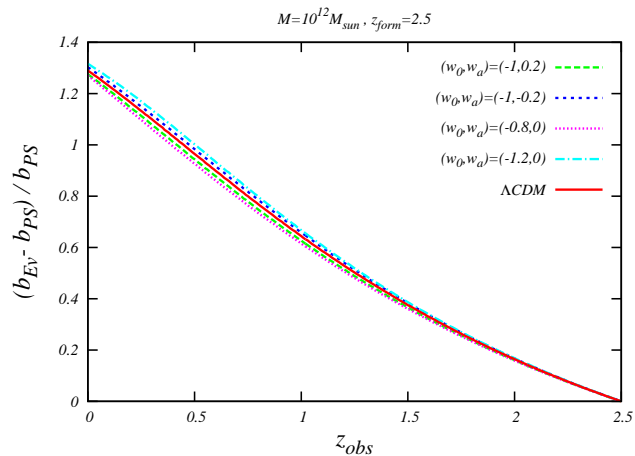}
\caption{The ratio of Evolved bias to Press Schechter bias minus one is plotted versus the observation redshift for different models of dark energy. The red solid line indicates the $\Lambda$CDM, the green long-dashed line shows a dark energy model with $(w_0,w_a)=(-1,0.2)$ in CPL parameterizations. The blue dashed line is for a model with $(w_0,w_a)=(-1,-0.2)$. The magenta dotted line  is for equation of state $(w_0,w_a)=(-0.8,0)$ and cyan dashed-dotted line indicate $(w_0,w_a)=(-1.2,0)$. The dark matter halo is assumed to have a mass of $M=10^{12} M_{\odot}$ as a typical mass of a SDSS galaxy catalog galaxies host with formation redshift $z_{f}=2.5$.}
\label{fig:bias-g}
\end{figure}

In principle, the velocity difference can be position-dependent. By assuming that matter and galaxies experience the same gravitational potential we find that $\textbf{v}_g = \textbf{v}_m$. In other words, even if there was a bias between $\textbf{v}_g$ and $\textbf{v}_m$ initially, it will be washed away eventually.
Following the same logic, The difference of density contrasts of galaxy and bias is written as
\begin{equation}\label{eq:deltaprime}
(\delta_g - \delta_m)^\prime = -\nabla . (\textbf{v}_g - \textbf{v}_m) \equiv - \frac{1}{a}\nabla . \textbf{u}_0,
\end{equation}
where $\textbf{u}_0$ is a function of position alone, not depending on time. The integration of Eq.(\ref{eq:deltaprime}) gives
\begin{equation}
\delta_g - \delta_m = -\nabla . \textbf{u}_0\int \frac{da}{a^3 H} + \Delta_i.
\end{equation}
The first term in the rhs is a decaying term, so it has no importance in our discussion, whatsoever. While the second term $\Delta_i$ plays a crucial role ($i$ indicates to the initial time). So the bias parameter will evolve as
\begin{equation}
b(a)\equiv \frac{\delta_g (a)}{\delta_m (a)}=1 + \frac{\Delta_i }{\delta_m (a)}.
\end{equation}
As mentioned before, $\Delta_i$ is time-independent. So we can write it as follow
\begin{equation}
\Delta_i =(b_i -1)\delta_m (a_i ),
\end{equation}
where $b_i$ and $\delta_m$ have been evaluated at some initial time. Now, assuming that the galaxy bias is initially scale-dependent we can show that  in Fourier space
\begin{equation}
b(a,k)=1+(b_i -1)\frac{\delta_m (a_i , \textbf{k})}{\delta_m (a , \textbf{k})}.
\end{equation}
Now we can exchange the ratio of the density contrasts with growth function $D(k,z)$, which in cosmological models deviating from $\Lambda$CDM can be scale-dependent, as well.
Now the bias in observed redshift $z_{obs}$ can be related to the Lagrangian bias computed in the formation redshift $z_{f}$ as follows (the initial time is set equal to the formation time):
\begin{equation}\label{eq:evolved}
b_{Ev}(k_0 , z_{obs}; M, z_{f})=1 + [b_{PS}(M, z_{f})-1]\frac{D(k_0 ,z_{f})}{D(k_0 ,z_{obs})},
\end{equation}
where $b_{Ev}$ is the evolved bias and $b_{PS}$ is the Press-Schechter bias in formation time.
The point here is that, if we want to use the cosmological observations which are based on the biased tracers of dark matter (like ISW-galaxy cross correlation function), we have to redefine the evolved bias parameter due to the dynamics of the new cosmological model.
For example in Fig.(\ref{fig:bias-LRG}), we plot the ratio of the evolved bias to Press-Schechter bias minus one, in terms of the redshift of observation. In order to find the Press-Schechter bias, we assume that the LRGs are typically hosted by dark matter halos of $M \sim 10^{13}M_{\odot}$. Another piece of information we need is the formation redshift of the host halos. We assume that the halo of LRGs are approximately formed in redshifts where their mass variance becomes unity $\sigma(M,z_{f})\simeq 1$. This condition sets the formation time equal to $z_{f}\simeq 2$ for this sample. This very rough approximation is in good agreement with simulation data \cite{Masaki:2012gh}. Fig.(\ref{fig:bias-LRG}) shows that the ratio becomes larger in low redshifts. In Fig.(\ref{fig:bias-g}), we plot the same ratio of the evolved bias and Press-Schechter bias for SDSS, galaxy sub-sample with r-band magnitude $18 < r < 21$ (We will discuss about the galaxy sub-samples in more detail in next section). We assume that host of this galaxies have mass $M \sim10^{12}M_{\odot}$. According to our estimate the formation redshift of this halos will be $z_{f}\simeq 2.5$. It is obvious that in the redshift of the formation, the evolved bias is the same as the Press-Schechter bias.
In the next section we will show the theoretical result for ISW-galaxy correlation with emphasizing on the degeneracy of free parameter of bias model and DE clustering.


\section{Galaxy catalog, bias and ISW-g correlation}
\label{Sec:Res}

In this section we compare and show the degeneracy of DE models  prediction for the cross correlation of ISW and galaxy angular power spectrum with bias parameter. We also discuss the importance of the {\it{joint kernel}}, introduced in this work, to deduce the effect of a DE model on the signal.
It is worth to mention that the idea of evolved bias via a phenomenological parametrization is discussed in Xia et al. \cite{Xia:2009dr} for quasars data sample, and in Ferraro et al. \cite{Ferraro:2014msa} for Wide field Infrared Survey explorer (WISE) selected galaxies \cite{Yan:2012yk}. Ade et al. \cite{Ade:2015dva} (Planck 2015 results XXI) study this problem with radio sources from the NVSS catalogue, galaxies from the optical SDSS data set, and the infrared WISE survey.
Another point is that because of low signal to noise, the ISW-LSS cross correlation data is not capable of constraining too many cosmological parameters alone. That is why we need complementary observations beside the ISW-LSS data to pin down the DE models.
In this section, we will discuss the effect of the evolved bias and clustering of DE on ISW-g signal, systematically.
In the first subsection we start with case study. We study different examples with different amounts of the equation of state of DE and the bias function, in which the physical interpretation of the signal due to chosen parameters is discussed. In subsection B, first we find the best parameter fit for equation of state and the bias function considering phenomenological parameterizations. In order to study the joint fitting function of bias and DE parameters we use the galaxy-galaxy autocorrelation data, as well.
The data samples in ISW-g cross correlation chosen here, are from SDSS galaxy sub-sample and LRGs, which are correlated with the CMB map of WMAP third year data \cite{Giannantonio:2008zi}. It is obvious that the ISW-galaxy signal depends on the background cosmological parameters, therefore it is worth to repeat the analysis presented in the upcoming two subsections with Planck-LSS correlation data and the cosmological parameters driven from Planck 2015 \cite{Ade:2015xua}.

\footnotetext[1]{The matter density parameter from Planck data $\Omega_m\simeq 0.31$ is higher than the WMAP, and the Hubble parameter of Planck $h\simeq 0.67$ is smaller than WMAP. This will change the Hubble parameter, comoving distance, growth function, the derivative of the growth function and matter power spectrum. All of these functions appeared in Eq.(\ref{eq:cgt}) and can change the signal of $C(\theta)$ (depending on the angle of separation) about $10\%$ on average.}

\subsection{Case Study}
\begin{figure}
\centering
\includegraphics[width=0.5\textwidth]{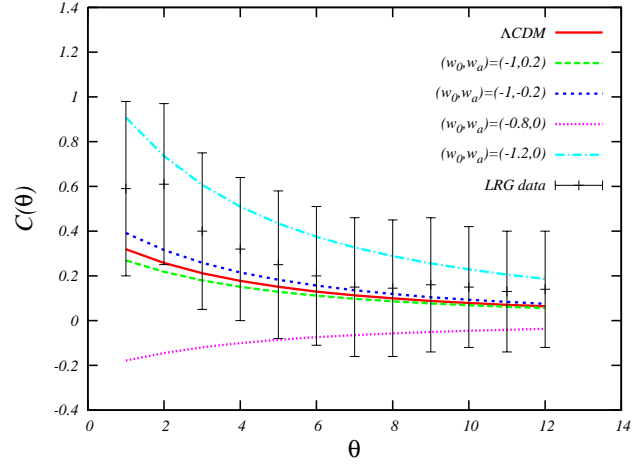}
\caption{ The angular power spectrum of the ISW-galaxy in micro-Kelvin is plotted versus the angle of separation,  for dark energy models. The red solid line indicates the $\Lambda$CDM, the green long-dashed line shows a dark energy model with $(w_0,w_a)=(-1,0.2)$. The blue dashed line is for a model with $(w_0,w_a)=(-1,-0.2)$, the magenta dotted line is for equation of state $(w_0,w_a)=(-0.8,0)$ and cyan dashed-dotted line indicates $(w_0,w_a)=(-1.2,0)$. The bias parameter is set to a constant ($b=1.8$). The data points are taken from Luminous Red Galaxy sample.}
\label{fig:ClLRG}
\end{figure}

In this subsection we study the angular power spectrum of ISW-galaxy cross correlation for different models of clustered DE, considering the corresponding bias from evolving scheme.
In Fig.(\ref{fig:ClLRG}), the ISW-galaxy angular correlation function is plotted versus separation angle for $\Lambda$CDM model (red solid line) in comparison with the DE model assuming the CPL parametrization introduced in Eq.(\ref{eq:CPL}).
We study the deviation from  $\Lambda$CDM in Fig.(\ref{fig:Q}), using the combination of four points from the parameter space of equation of state $(w_0,w_a)=(-1,0.2)$, $(w_0,w_a)=(-1,-0.2)$, $(w_0,w_a)=(-1.2,0)$ and $(w_0,w_a)=(-0.8,0)$. The first three points are compatible with the Planck data (angular correlation of temperature) in combination with Baryon Acoustic Oscillation (BAO) data and  SNeI data (The Joint Light-curve Analysis sample (JLA) \cite{Betoule:2014frx}, and local measurement of $H_0$ in $2\sigma$ confidence \cite{Ade:2015xua}). The last point $(w_0,w_a)=(-0.8,0)$ has a  $2\sigma$ tension with the Planck data. The LRG data sample of ISW-g correlation function shows that $(w_0,w_a)=(-0.8,0)$ is already ruled out with the constant bias parameter. This means that the constraints from ISW-g cross correlation is in the same direction as the other cosmological probes and if we can determine the bias parameter well, we can use ISW-g cross correlation data to tighten the constraints. The effect of changing the bias will be studied later in this section.

\begin{figure}
\centering
\includegraphics[width=0.5\textwidth]{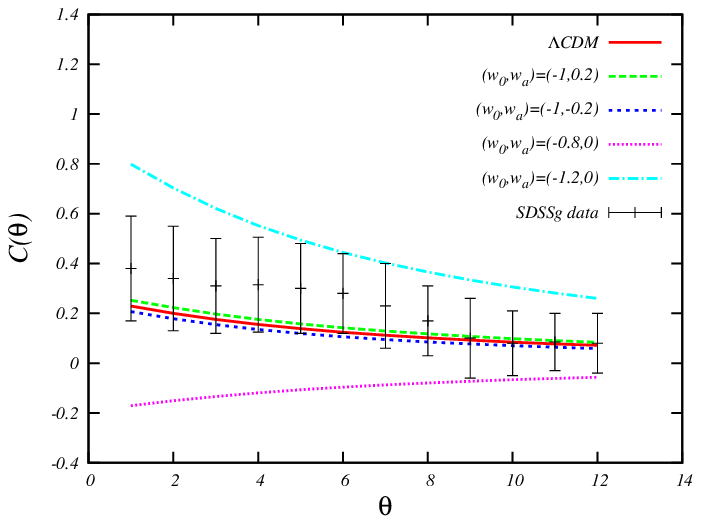}
\caption{The angular power of the ISW-galaxy in micro-kelvin is plotted versus the angle for dark energy models. The red solid line indicates the $\Lambda$CDM, the green long-dashed line shows a dark energy model with $(w_0,w_a)=(-1,0.2)$. The blue dashed line is for a model with $(w_0,w_a)=(-1,-0.2)$, the magenta dotted line is for equation of state $(w_0,w_a)=(-0.8,0)$ and cyan dashed-dotted line indicates $(w_0,w_a)=(-1.2,0)$. The bias parameter is set to a constant ($b=1.0$). The data points are taken from SDSS Galaxy sample.}
\label{fig:ClSDSS}
\end{figure}


\begin{figure}
\centering
\includegraphics[width=0.5\textwidth]{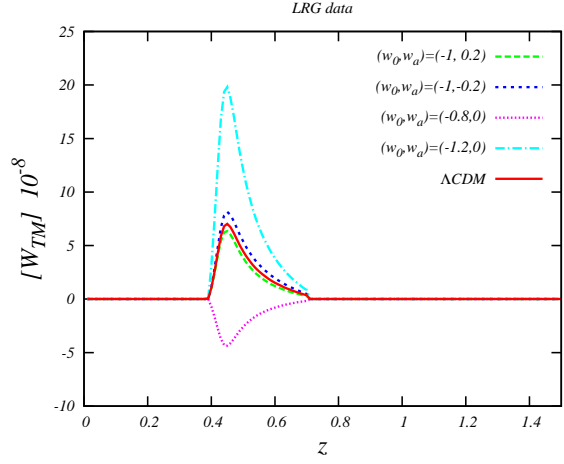}
\caption{The joint Kernel of ISW-galaxy $W_{TM}$ for the Luminous Red Galaxies (LRGs) survey is plotted versus redshift.
The red solid line indicate the $\Lambda$CDM, the green long-dashed line shows a dark energy model with $(w_0,w_a)=(-1,0.2)$. The blue dashed line is for a model with $(w_0,w_a)=(-1,-0.2)$, the magenta dotted line is for equation of state $(w_0,w_a)=(-0.8,0)$ and cyan dashed-dotted line indicate $(w_0,w_a)=(-1.2,0)$.}
\label{fig:WTM-LRG}
\end{figure}

\begin{figure}
\centering
\includegraphics[width=0.5\textwidth]{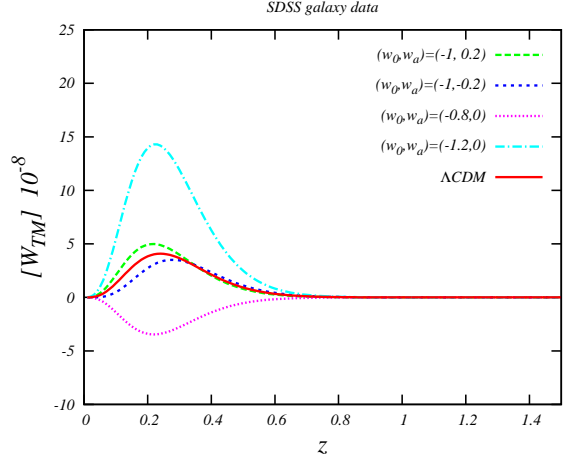}
\caption{The joint Kernel of ISW-galaxy $W_{TM}$ for a sub-sample of galaxies from SDSS sub-sample  is plotted versus redshift. The red solid line indicate the $\Lambda$CDM, the green long-dashed line shows a dark energy model with $(w_0,w_a)=(-1,0.2)$. The blue dashed line is for a model with $(w_0,w_a)=(-1,-0.2)$, the magenta dotted line is for equation of state $(w_0,w_a)=(-0.8,0)$ and cyan dashed-dotted line indicate $(w_0,w_a)=(-1.2,0)$.}
\label{fig:WTM-g}
\end{figure}

\begin{figure}
\centering
\includegraphics[width=0.5\textwidth]{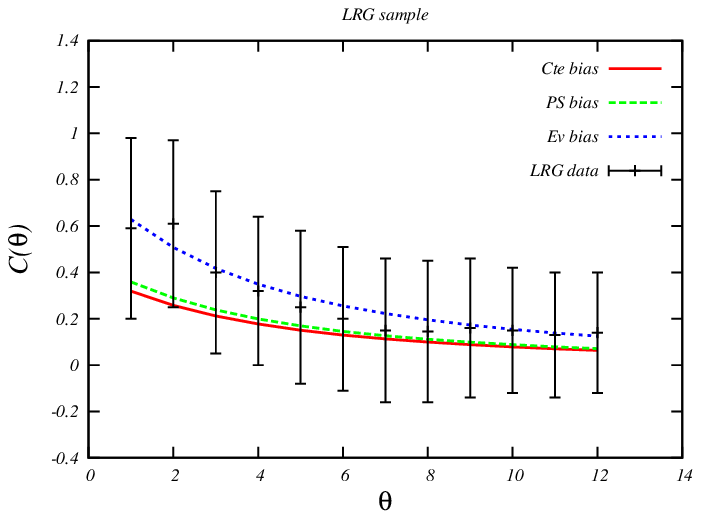}
\caption{The angular power of the ISW-galaxy in micro-kelvin is plotted versus the angle for $\Lambda$CDM model with different biases.
The red solid line showed the constant bias ($b=1.8$), the green long dashed line showed the Press-Schechter bias for dark matter halo mass of $M=10^{13}M_{\odot}$ and the blue dashed line shows the evolved bias with the assumption of $z_{f}=2$. The data points are taken from LRG sample.}
\label{fig:ClbiasLRG}
\end{figure}

\begin{figure}
\centering
\includegraphics[width=0.5\textwidth]{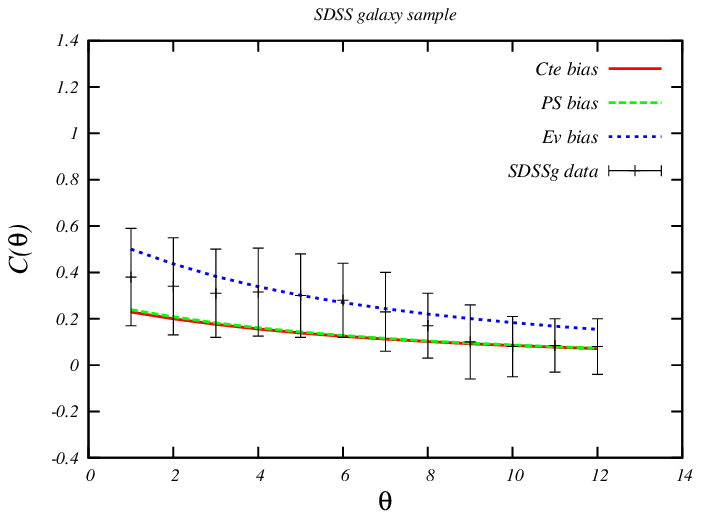}
\caption{The angular power of the ISW-galaxy in micro-kelvin is plotted versus the angle for $\Lambda$CDM model with different biases.
The red solid line shows the constant bias ($b=1.0$), the green long dashed line shows the Press-Schechter bias for dark matter halo mass of $M=10^{12}M_{\odot}$ and the blue dashed line shows the evolved bias with assumption of $z_{f}=2.5$. The data points are taken from SDSS sample.}
\label{fig:ClbiasSDSS}
\end{figure}

\begin{figure}
\centering
\includegraphics[width=0.5\textwidth]{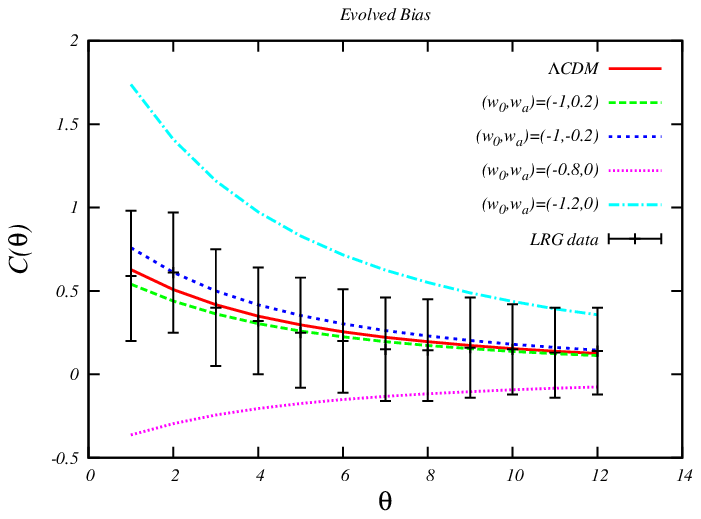}
\caption{The angular power of the ISW-galaxy in micro-kelvin is plotted versus the angle of separation, for dark energy models. The red solid line indicates the $\Lambda$CDM, the green long-dashed line shows a dark energy model with $(w_0,w_a)=(-1,0.2)$. The blue dashed line is for a model with $(w_0,w_a)=(-1,-0.2)$, the magenta dotted line is for equation of state $(w_0,w_a)=(-0.8,0)$ and cyan dashed-dotted line indicate $(w_0,w_a)=(-1.2,0)$. The bias is the evolved one for dark matter halos of $M=10^{13}M_{\odot}$ and $z_{f}=2.0$. The data points are taken from SDSS LRG sample.}
\label{fig:ClDELRG}
\end{figure}
\begin{figure}
\centering
\includegraphics[width=0.5\textwidth]{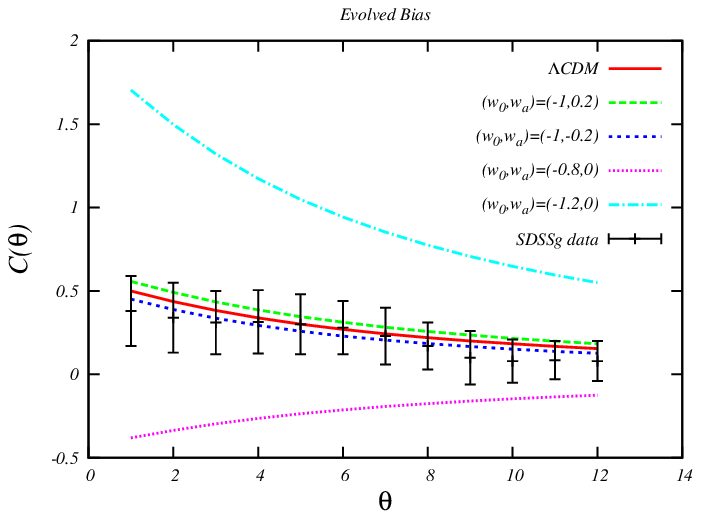}
\caption{The angular power of the ISW-galaxy in micro-kelvin is plotted versus the angle of separation, for dark energy models. The red solid line indicates the $\Lambda$CDM, the green long-dashed line shows a dark energy model with $(w_0,w_a)=(-1,0.2)$. The blue dashed line is for a model with $(w_0,w_a)=(-1,-0.2)$, the magenta dotted line is for equation of state $(w_0,w_a)=(-0.8,0)$ and cyan dashed-dotted line indicate $(w_0,w_a)=(-1.2,0)$. The bias is the evolved one for dark matter halos of $M=10^{12}M_{\odot}$ and $z_{f}=2.5$. The data points are taken from SDSS Galaxy sample.}
\label{fig:ClDESDSS}
\end{figure}

The data points in Fig. (\ref{fig:ClLRG}), are from the cross correlation of the CMB data with  LRGs which are extracted from the SDSS catalog \cite{Blake:2006kv}. LRGs are suitable choice as a tracer of dark matter distribution as they have a deeper redshift distribution (with a mean redshift of $z\sim 0.5$) than ordinary
galaxies. Thus they have been used to find evidence for the ISW effect, previously \cite{Scranton:2003in,Padmanabhan:2004fy}.
In this analysis we use the data points from \cite{Giannantonio:2008zi} which is processed from "MegaZ LRG" sample \cite{Blake:2006kv,Collister:2006qg} and contains 1.5 million objects from the SDSS DR6 selected with a neural network.
Theoretical curves in Fig.(\ref{fig:ClLRG}) are plotted with constant halo dark matter bias ($b = 1.8$) \cite{Giannantonio:2008zi}. The error bars in the LRG catalog are quite large in $1\sigma$ level. Accordingly,  all the data points chosen from free parameter space of CPL parameterizations except $(w_0,w_a)=(-0.8,0)$ are compatible with ISW-galaxy data in the proposed level of confidence.\\
In Fig.(\ref{fig:ClSDSS}), the same theoretical models are plotted in comparison to the standard $\Lambda$CDM prediction. In this figure, we compare the theoretical curves with another catalog of galaxies taken from SDSS galaxy sample \cite{Giannantonio:2008zi}. The data set is from SDSS Sixth Data Release (DR6) \cite{AdelmanMcCarthy:2007aa,York:2000gk}. From this catalog a magnitude limited sub-sample is chosen as $18 < r < 21$, where  ``r is the extinction corrected SDSS calibrated model magnitude" \cite{Giannantonio:2008zi}. In this case, the catalog contains almost $3\times 10 ^6$ galaxies.
Using the chosen SDSS galaxy sub-sample as a probe of ISW-g correlation, we find a more significant tension between DE model with the equation of state $(w_0,w_a)=(-0.8 , 0 )$ and $(w_0,w_a)=(-1.2,0)$.
In future wider and deeper galaxy surveys, we will be able to constrain the DE equation of state more precisely. This is because the ISW-galaxy correlation is a dynamical probe of DE models. A probe which in the same time captures the effect of the deviation from $\Lambda$CDM in background (Hubble parameter) and in the level of perturbations via matter power spectrum and growth function.\\
Important point here is that there is not a one to one relation between the behavior of the deviation from $\Lambda$CDM shown in $Q$ parameter (plotted in Fig.(\ref{fig:Q})) and the signal of ISW-galaxy correlation function. This can be anticipated because the ISW-galaxy signal is not sourced only by the clustered DE model but it has the effect of the dark matter distribution on it as well. In order to understand this effect, in Figs.(\ref{fig:WTM-LRG}) and (\ref{fig:WTM-g}), we showed the evolution of  joint kernel of ISW-galaxy ($W_{TM}$) versus redshift for different models of DE. The joint kernel is defined as the multiplication of two kernels $W_M$ and $W_T$ as
\be
W_{TM}= W_T\times W_M,
\ee
where $W_M$ is defined as below
\be
W_{M}=\frac{dN}{dz}\tilde{D}(z),
\ee
in which $dN/dz$ is the galaxy distribution function defined previously, and $~\tilde{}~$ indicates that the growth function in a specific DE model could be different from the standard case of $\Lambda$CDM. The $W_T$ is defined as
\be
W_T=3k^{-2}({\cal{H}}\Omega_m \tilde{D}(z)Q)_{,\eta},
\ee
where $, \eta$ is the derivative with respect to conformal time.
Now by comparing the Fig.(\ref{fig:ClLRG}) with Fig.(\ref{fig:WTM-LRG}) and Fig.(\ref{fig:ClSDSS}) with Fig.(\ref{fig:WTM-g}) we will find, a one to one relation between $W_{TM}$ and the ISW-g correction function results. The distribution function of galaxies in two catalogs are so crucial that it changes the amplitude of the signal for two DE models with $(w_0,w_a)=(-1,0.2)$ and $(w_0,w_a)=(-1,-0.2)$ for the SDSS- galaxy sample.
This will have an important implication in choosing the multi-galaxy sub-sample for probing DE in future observations, because each galaxy sample probes  a specific redshift interval more effectively.
This shows that except the evolution of DE model, it is important to understand the distribution of the dark matter tracers.
In what follows we will study the effect of dark matter halo bias on the ISW-galaxy signal.
In Fig.(\ref{fig:ClbiasLRG}), we plot the angular power of ISW-g effect for $\Lambda$CDM with different biases versus angle of separation. The red solid line shows the result using the constant bias $b=1.8$. The Press Schechter bias (green long dashed line) is obtained by the assumption of $M=10^{13}M_{\odot}$ for the mass of the dark matter halo which is the host of LRG. The ISW-g signal which is obtained by Press-Schechter is very similar to the one obtained from constant bias. The blue dashed line shows the ISW-g signal with evolved bias obtained from Eq.(\ref{eq:evolved}) with assumption of $z_{f}=2.0$ \cite{Masaki:2012gh}. The $\Lambda$CDM fits with evolved bias improved statistically by $\Delta\chi^2 \simeq 2.1$ in comparison with the signal obtained from constant bias.\\
It is interesting to note that there are studies \cite{Giannantonio:2008zi,Ho:2008bz,Goto:2012yc} showing the data are $\sim1\sigma$ to $\sim2\sigma$ higher than what we expect from the standard model, the same tension is also obtained from stacking of the voids and clusters  \cite{Granett:2008ju,Ade:2013dsi}. Our results show that the evolved bias can relax this tension, which is the same conclusion in \cite{Ferraro:2014msa}, where the lensing data is used to constrain the bias parameter. It is shown that the bias is not constant in all the range of integration of ISW-g signal. \\
In Fig. (\ref{fig:ClbiasSDSS}), we plot the angular power of ISW-g effect for $\Lambda$CDM with different biases versus angle of separation, this time for the SDSS galaxy sample. The red solid line shows the result using the constant bias $b=1.0$. The Press-Schechter bias (green long dashed line) is obtained by the assumption of $M=10^{12}M_{\odot}$ as the Mass of host dark matter of luminous red galaxies. The ISW-g signal which is obtained by Press-Schechter is very similar to the effect of constant bias here, as well. The blue dashed line shows the ISW-g signal with evolved bias obtained from Eq.(\ref{eq:evolved}) with assumption of $z_{f}=2.5$ \cite{vanDokkum:2013hza}. The $\Lambda$CDM fit with evolved bias, improves the fit by $\Delta\chi^2 \simeq 1.9$. Once more this result shows that the evolved bias could be a probable solution to the discrepancy of $\Lambda$CDM and  observational data. \\
Now we come back to study of the  DE clustering with considering the evolved bias term.
In Fig.(\ref{fig:ClDELRG}) and Fig.(\ref{fig:ClDESDSS}), we study the effect of the evolved bias on the signal of ISW-g cross correlation. As we anticipate from Fig.(\ref{fig:bias-LRG}) and Fig.(\ref{fig:bias-g}), the evolved bias is almost twice the Press-Schechter bias in intermediate redshifts where two catalogs of LRG and SDSS galaxies are most effective. Thus taking into account the evolved bias parameter, the equation of state parameters ($w_0$,$w_a$) can be constrained  more tightly. It should be noted that changing the background cosmology parameters change the evolved bias through the Eq.(\ref{eq:evolved}). This study has a very promising conclusion, obtaining more accurate data of ISW galaxy cross correlation in the future plus knowledge of the more realistic bias parameter (by theoretical considerations and independent observations), will help us to pin down the equation of state of DE. In the next subsection we study the parameter space of DE models and bias with likelihood analysis and we will show how the evolved bias tighten the constraint on clustered DE models.

\subsection{Parameter Estimation}
In this subsection we constrain the free parameters of DE and bias models simultaneously. This is done by using the ISW-galaxy correlation data obtained from two previously mentioned catalogs (SDSS galaxies and LRG data) and also the autocorrelation of galaxy-galaxy data of the same two samples \cite{Giannantonio:2008zi}.
We should note that the signal to noise of the ISW-galaxy is low and therefore we need complementary observations to constrain the cosmological model, and this can not be done only by ISW-g data alone. Accordingly we will use the autocorrelation data as a complimentary probe to constrain the free parameters.

In order to do the parameter estimation systematically, first we assume that the physics of the bias parameter is fixed by other observations or a known theory, and we use the ISW-g data to constrain the DE parameters. Afterwards we relax this assumption and constrain the DE and bias parameters simultaneously, considering the bias model parameters as new degrees of freedom.

\begin{figure}[!htb]
\minipage[t]{0.45\linewidth}
\advance\leftskip-1.5cm
\textbf{SDSS galaxy and LRG sample , cte bias}\par\medskip
\includegraphics[width=6.6cm, height=5.0cm]{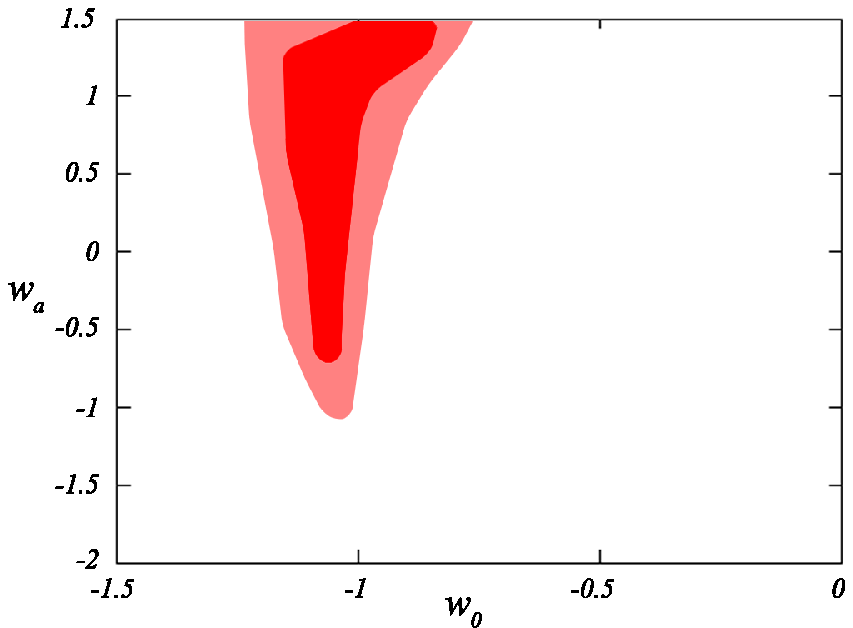}
\caption{$1\sigma$ and $2\sigma$ confidence levels of free parameters of dark energy model ($w_0$ and $w_a$) using the observational data of ISW-galaxy cross correlation. The sample contains both samples of SDSS-galaxy and LRG catalog. The bias is assumed to be constant for each sample.}\label{fig:w0wa-bcte}
\endminipage\hfill
\minipage[t]{0.45\linewidth}
\advance\leftskip-1.5cm
\textbf{SDSS galaxy and LRG sample , Ev. bias}\par\medskip
\includegraphics[width=6.6cm, height=5.0cm]{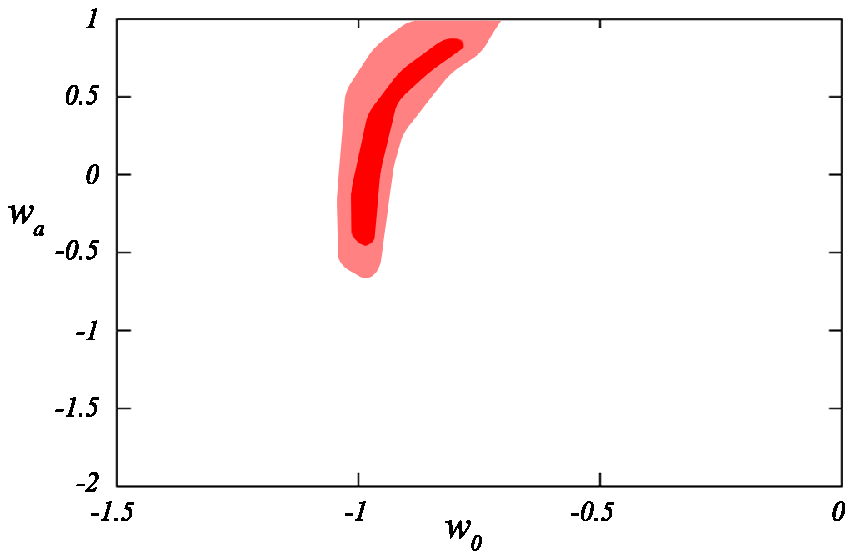}
\caption{$1\sigma$ and $2\sigma$ confidence levels of free parameters of dark energy model ($w_0$ and $w_a$) using the ISW-galaxy cross correlation.  The data contains both samples of SDSS-galaxy and LRG catalog. We consider the evolved bias scheme as described in Sec.(\ref{Sec:Bias}) with a formation redshift and mass of the host halo fixed for each sample separately.}\label{fig:w0wa-bevo}
\endminipage\hfill
\end{figure}

In Fig.(\ref{fig:w0wa-bcte}), we plot the $1\sigma$ and $2\sigma$ levels of confidence for free parameters of the clustering DE ($w_0$ and $w_a$). Note that we use the CPL parametrization as in previous subsection, $w=w_0+w_a(\frac{z}{1+z})$, with the assumption of barotropic clustered DE fluid.
For this plot, we only use ISW-g data, however both samples of data (SDSS and LRG) are used in $\chi^2$ fitting where we assume a fixed constant bias for each data catalog. The bias for SDSS-galaxy catalog is set to $b_{(sdss)} = 1 $ and the bias parameter for LRG is set to $b_{(lrg)}=1.8$.
In following we will relax the assumption of fixed bias and will take the constant bias as a free parameter.
In Fig.(\ref{fig:w0wa-bevo}), we plot the confidence level for free parameters of DE, where we assumed the evolved bias model,  instead of the constant one. For each data sample the bias is calculated by the procedure introduced in Sec.(\ref{Sec:Bias}). As a reminder, for SDSS galaxies, the bias is calculated with the formation redshift set to $z_{f}\sim 2.5$ and we also assign the host halo of $M=10^{12}M_{\odot}$ to each galaxy.
For LRG sample, we set $z_{f}\sim 2.0$ and the host halo mass to $M=10^{13}M_{\odot}$. Fig.(\ref{fig:w0wa-bcte}) and Fig.(\ref{fig:w0wa-bevo}) show the consistency of our conclusions stated in previous subsection. From parameter estimation we show that the DE fluid with equation of state of $(w_0,w_a)=(-1.2,0)$ and $(w_0,w_a)=(-0.8,0)$  are incompatible with data. These two figures, also show that for the evolved bias case, the constraints on DE parameters are tighter. This is a reasonable conclusion, because the evolved bias term, does not introduce a new parameter for fitting and accordingly does not loosen the constraints on the equation of state. In contrary, the evolved bias model introduces a more complex theoretical function of ISW-galaxy cross correlation curve, which must be fitted with the observed data. Therefore, the fitting procedure becomes more difficult and consequently the confidence levels shrink. \\
One step further, in order to perform more comprehensible  and coherent statistical analysis, we study the whole parameter space, i.e., looking for constraints on DE parameters and bias, jointly. For this task, we should note that the ISW-g cross correlation data has a low signal to noise. Accordingly, to compensate this problem we use the galaxy-galaxy autocorrelation data as well \cite{Giannantonio:2008zi}.
In first stage of joint analysis, we assume a constant bias model (which is assumed to be a free parameter) and a phenomenological CPL parametrization of clustered DE.

\begin{figure}
\centering
\includegraphics[width=0.8\textwidth]{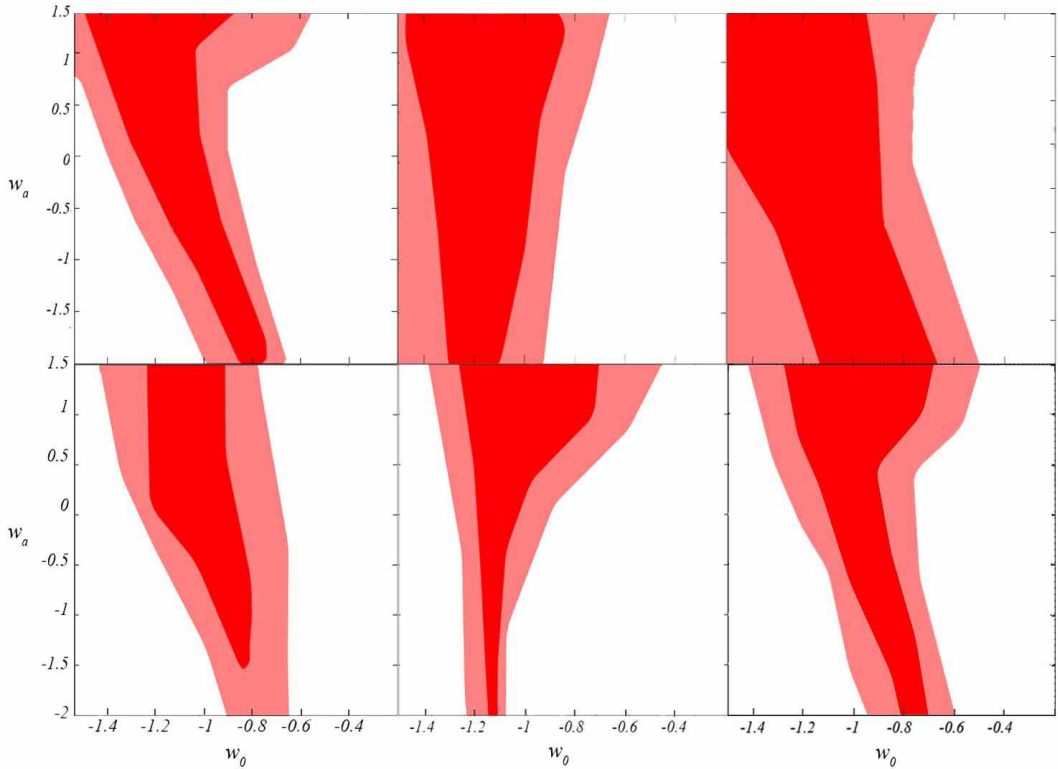}
\caption{The $1\sigma$ and $2\sigma$ confidence level of free parameters of dark energy model is plotted. The top row figures are for a constant bias model.  The bottom row is for evolved bias model. The left panels are confidence level obtained by using the LRG data of galaxy-galaxy autocorrelation and the cross correlation of ISW-galaxy. The middle panels used the SDSS-galaxy data and the right panels used both catalog data. Note that the right panel contour plots are obtained by marginalizing over two (top-right:constant bias model) and four parameters (bottom right: evolved bias model). }
\label{fig:w0wa34}
\end{figure}

In Fig.(\ref{fig:w0wa34}), we plot the $1\sigma$ and $2\sigma$ confidence levels for the free parameters of DE model ($w_0$ and $w_a$) using the ISW-g correlation function data and the galaxy-galaxy autocorrelation function for LRG galaxy sample, SDSS-galaxy sample and both data. In the top panels, we investigate the constant bias model, where the bias is a free parameter and it is marginalized to obtain the confidence level on DE parameters.
In the top row of Fig.(\ref{fig:w0wa34}),  the left panel, shows the confidence level on DE parameters by using just the LRG sample. (Three free parameters $w_0$, $w_a$ and $b_{(lrg)}$, where we marginalized over $b_{(lrg)}$). In the middle panel we use the SDSS-galaxy data,(three free parameters $w_0$, $w_a$ and $b_{(sdss)}$, where we marginalized over $b_{(sdss)}$) and in the right panel we use both catalogs (four free parameters $w_0$, $w_a$, $b_{(lrg)}$ and $b_{(sdss)}$, where we marginalized over $b_{(lrg)}$ and $b_{(sdss)}$).
In the bottom row of Fig.(\ref{fig:w0wa34}), we investigate the evolved bias model. The left panel shows the confidence level on DE parameters by using just the LRG sample (four free parameters $w_0$, $w_a$, $z_{f(lrg)}$ and $M_{(lrg)}$, where we marginalized over two of them $z_{f(lrg)}$ and $M_{(lrg)}$). In the middle panel we use the SDSS-galaxy data ,(four free parameters $w_0$, $w_a$, $z_{f(sdss)}$ and $M_{(sdss)}$, where we marginalized over two of them $z_{f(sdss)}$ and $M_{(sdss)}$) and in the right panel we use both catalogs (six free parameters $w_0$, $w_a$, $z_{f(lrg)}$ , $M_{(lrg)}$, $z_{f(sdss)}$ and $M_{(sdss)}$, where we marginalized over all bias related parameters).\\
It is seen that by adding the bias model as a free parameter, the constrains on DE parameters are loosen, as we anticipate.
An important point to indicate as well is that the compatibility of the phenomenological model with equation of state $(w_0,w_a)=(-1.2,0)$  and $(w_0,w_a)=(-0.8,0)$, which are chosen in previous section to study the effect of DE with the knowledge of the bias model is changed when we assume that the best fit-parameters of bias model is unknown and must be fixed by data. The equation of state $(w_0,w_a)=(-1.2,0)$ with a constant bias model is in agreement with data. However, this equation of state is in $1\sigma$ tension with the assumption of  evolved bias model. The equation of state with $(w_0,w_a)=(-0.8,0)$ is in $2\sigma$ tension with the data, when we use the constant bias model and  each catalog separately. However in constant bias model with two catalogues the tension reduces to $1\sigma$ level. In the evolved bias case the LRG catalog and the joint catalogs,  shows $1\sigma$ tension. However when we use the SDSS catalog we have $2\sigma$ tension. We should note that, as discussed in previous section, the SDSS-g error bars are smaller and put more tighter constraints on parameters. This results shows that we can fix the bias model and its parameter from independent observations, we can constrain the DE parameters better.  Otherwise we will face with a considerable amount of degeneracy.
\begin{figure}[!htb]
\minipage[t]{0.45\linewidth}
\advance\leftskip-1.5cm
\includegraphics[width=0.8\textwidth]{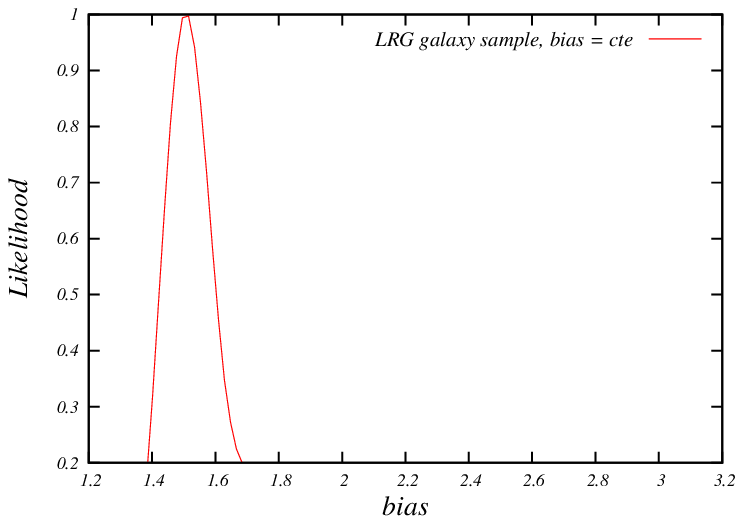}
\caption{The likelihood of a constant bias term using the ISW-galaxy correlation function and the galaxy-galaxy autocorrelation function using the LRG galaxy sample data . Dark energy free parameters ($w_0$ and $w_a$) are marginalized.}\label{fig:LikebLRG}
\endminipage\hfill
\minipage[t]{0.45\linewidth}
\advance\leftskip-1.5cm
\includegraphics[width=0.8\textwidth]{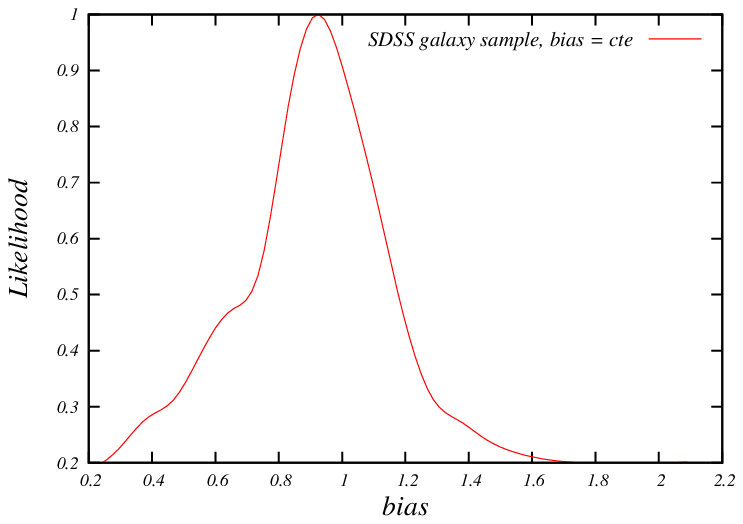}
\caption{The likelihood of a constant bias term using the ISW-galaxy correlation function and the galaxy-galaxy autocorrelation function using the SDSS galaxy sample data . Dark energy free parameters ($w_0$ and $w_a$) are marginalized.}\label{fig:LikebSDSS}
\endminipage\hfill
\end{figure} \\
In Fig.(\ref{fig:LikebLRG}), we plot the likelihood of constant bias parameter by ISW-g cross correlation and galaxy-galaxy auto-correlation data. The data set is LRG galaxy data sample. The ``constant bias" marginalized over DE parameters, has a value of $b_{lrg}=1.5 \pm 0.2$ in $1\sigma$ confidence level, which is in $1\sigma$ tension with the bias parameter fixed by galaxy autocorrelation function in $\Lambda$CDM model \cite{Giannantonio:2008zi}.
This tension is an indication that there is a degeneracy with DE parameters and dark matter halo bias. The constant bias obtained in \cite{Giannantonio:2008zi} is derived with the assumption of $\Lambda$CDM cosmology. Deviation from $\Lambda$CDM change the strength and scale dependence of matter fluctuations. Accordingly all observables like ISW-g cross correlation or galaxy-galaxy autocorrelation which are proportional to matter power spectrum are affected. The bias parameter is also changing the amplitude and scale dependence of matter power spectrum. We will plot the confidence level of parameter space of DE and bias, and we will show and discuss how this two class of parameters are degenerate.
In Fig.(\ref{fig:LikebSDSS}), we plot the likelihood of constant bias parameter for SDSS galaxy sample with joint analysis of ISW-g and galaxy-galaxy data.
The constant bias marginalized over DE parameters has a value of $b_{sdss}=0.9\pm 0.2$, which is in agreement with the bias parameter fixed by galaxy autocorrelation function in $\Lambda$CDM model\cite{Giannantonio:2008zi}. \\
The final part of the analysis in this section is focused on the evolved bias model which is introduced in detail in Sec.(\ref{Sec:Bias}). As we discussed, by knowledge of  mass of galaxy's host halo $M$ and its halo formation time $z_{f}$, we can obtain the bias parameter as introduced in Eq.(\ref{eq:evolved}). Accordingly, in evolved bias model, we have two free parameters $M$ and $z_{f}$. Thus we have to minimize a joint $\chi^2 = \chi^2 (w_0,w_a, z_{f}, M)$ with two data samples of ISW-g and galaxy-galaxy in hand.
\begin{figure}[!htb]
\minipage[t]{0.45\linewidth}
\advance\leftskip-1.5cm
\textbf{LRG galaxy sample, 4 parameter model fitting, bias}\par\medskip
\includegraphics[width=0.8\textwidth]{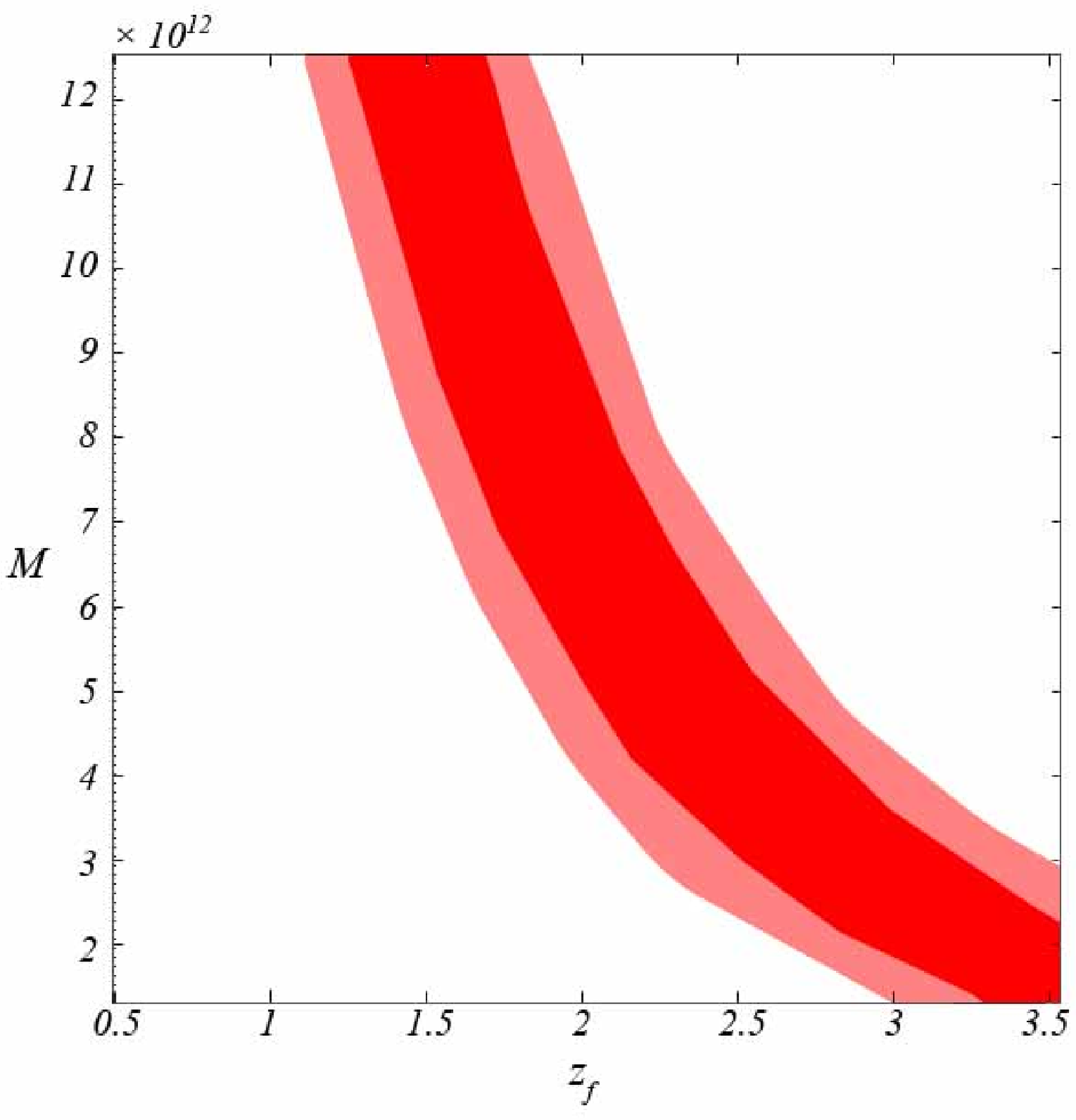}
\caption{Confidence levels of free parameters of bias model ($M$ and $z_{f}$) using the ISW-galaxy correlation function and the galaxy-galaxy autocorrelation function using the LRG galaxy sample data. The free parameters of Dark Energy Model ($w_0$ and $w_a$)  are marginalized.}\label{fig:LRG4b}
\endminipage\hfill
\minipage[t]{0.45\linewidth}
\advance\leftskip-1.5cm
\textbf{SDSS galaxy sample, 4 parameter model fitting, bias}\par\medskip
\includegraphics[width=0.8\textwidth]{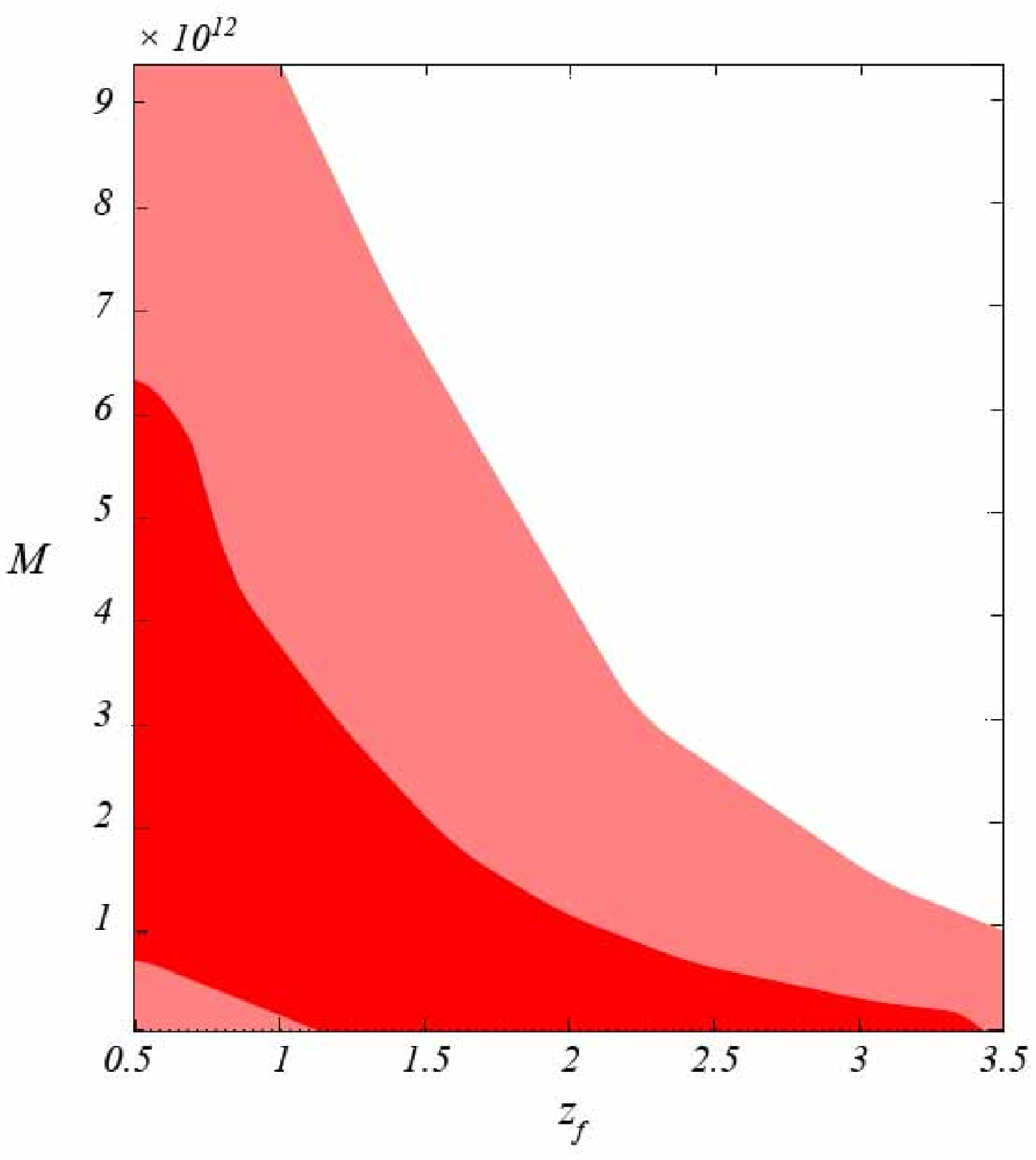}
\caption{Confidence levels of free parameters of bias model ($M$ and $z_{f}$) using the ISW-galaxy correlation function and the galaxy-galaxy autocorrelation function using the SDSS galaxy sample data . The free parameters of Dark Energy Model ($w_0$ and $w_a$)  are marginalized.}\label{fig:SDSS4b}
\endminipage\hfill
\end{figure}
Now it is possible to constrain the evolved bias parameters using the same data. In Fig.(\ref{fig:LRG4b}), we plot the $1\sigma$ and $2\sigma$ confidence levels of free parameters of bias model ($M_{(lrg)}$ and $z_{f(lrg)}$ using the ISW-galaxy correlation function and the galaxy-galaxy autocorrelation function for the LRG galaxy sample data. It shows the viable range of LRG halo mass and formation redshift. It is interesting to see that $M_{(lrg)}=10^{13}M_{\odot}$ and $z_{f}=2$ which is assumed through this work in previous sections, is consistent with our analysis in a $2\sigma$ level.
In Fig.(\ref{fig:SDSS4b}), we plot the $1\sigma$ and $2\sigma$ confidence levels of free parameters of bias model ($M$ and $z_{f}$) for SDSS galaxy sample. In the SDSS galaxy sample case, we find that our assumed values of $M_{(sdss)}=10^{12}M_{\odot}$ and $z_{f(sdss)}=2.5$ are consistent with the $\chi^2$ analysis again in just $2\sigma$ level. This is anticipated as we find the formation redshift via a very simple assumption that at the formation redshift the mass variance must be equal zero. This is criteria also is studied in standard $\Lambda$CDM model. When we change the background cosmology, the formation redshift is changed correspondingly.

\begin{figure}
\centering
\includegraphics[width=0.8\textwidth]{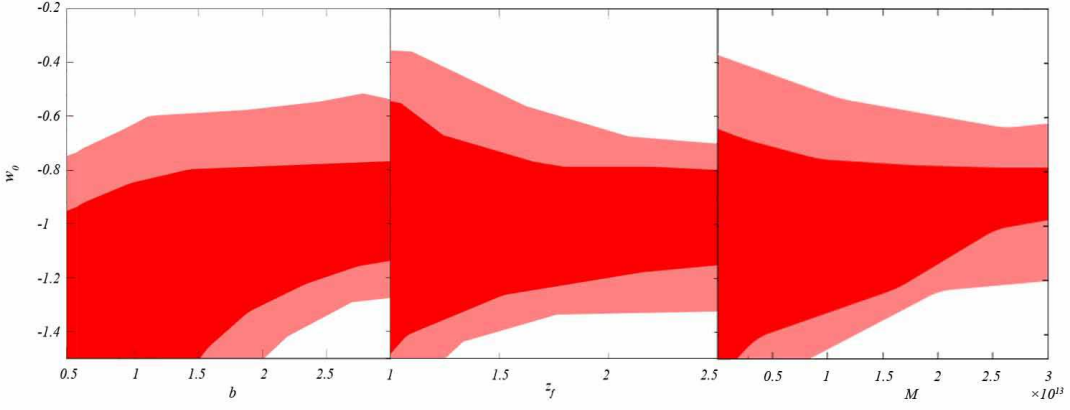}
\caption{The $1\sigma$ and $2\sigma$ confidence levels of $\omega_0$ and bias parameters. The left panel figure is confidence levels of $\omega_0$ and constant bias for three free parameter model. The middle and the right panel is for the confidence levels of $\omega_0$ with mass and redshift of the dark matter halo in four parameter model. The contour plots are obtained by using the LRG data.}
\label{fig:w0biasLRG}
\end{figure}


\begin{figure}
\centering
\includegraphics[width=0.8\textwidth]{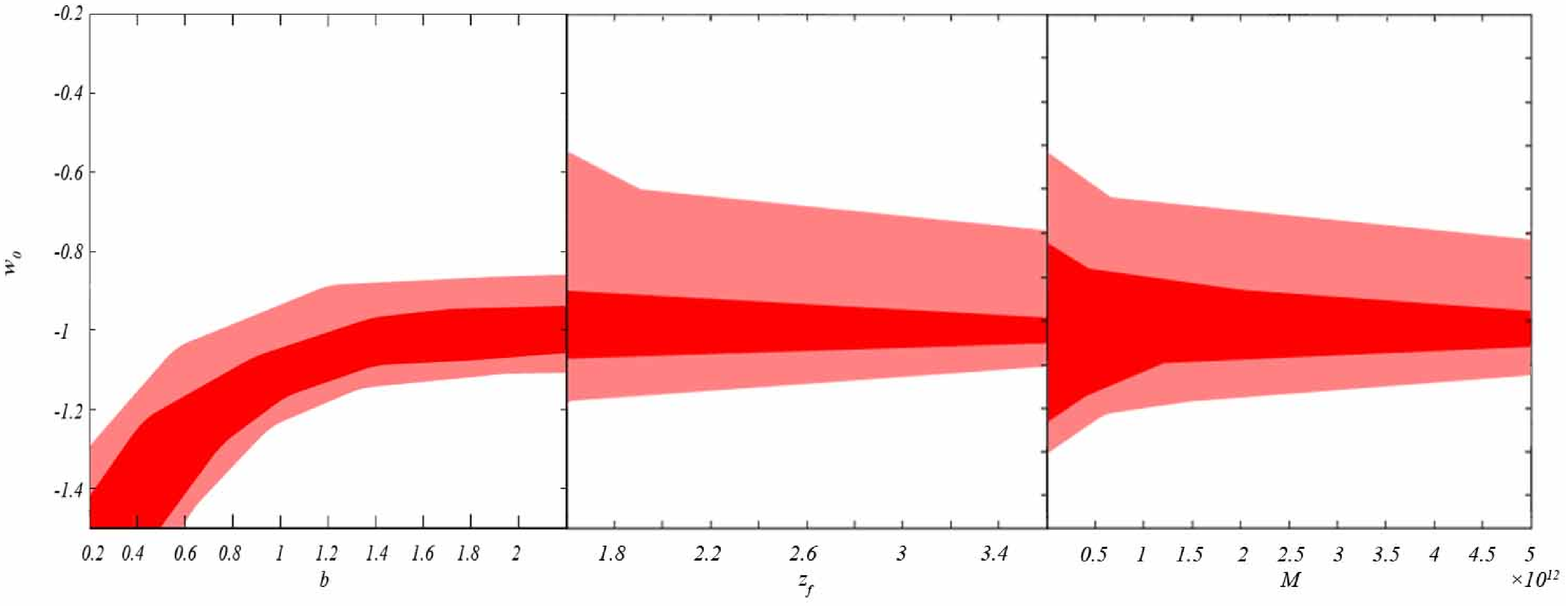}
\caption{The $1\sigma$ and $2\sigma$ confidence levels of $\omega_0$ and bias parameters. The left panel figure is confidence levels of $\omega_o$ and constant bias for three free parameter model. The middle and the right panel is for the confidence levels of $\omega_0$ with mass and redshift of the dark matter halo in four parameter model. The contour plots are obtained by using the SDSS-galaxy data.}
\label{fig:w0biasSDSS}
\end{figure}
As mentioned in the introduction, one of the main goals of this work is the study of the degeneracy of DE physics and the bias physics. In this direction,
in order to see these degeneracies better we plot the DE parameters versus bias parameter with their level of confidence. In Fig.(\ref{fig:w0biasLRG}) for the LRG catalog, we plot the confidence level of DE equation of state parameter $w_0$ versus constant bias in the left panel, formation redshift in the middle panel and the halo mass in the right panel. We should note that the left panel is for the constant bias model and the middle and the right one is for evolved bias model.
In Fig.(\ref{fig:w0biasSDSS}) the same configuration is plotted, this time for the SDSS catalog.
The constant bias has the most correlation with the $w_0$ parameter. This a figurative manifestation of DE and bias model degeneracy we discussed.
To wrap up all the results we list the best fit parameters in Table (I) which shows the best parameter of DE model and bias parameters.
\begin{table}
\centering
\begin{tabular}{|c|c|c|c|c|c|c|}
  \hline
  Model  & catalog & $w_0$  & $w_a$  & $b$  & $z_{f}$ & $M$    \\
 \hline
 \hline
  Constant bias  & LRG         & $-1.0 \pm 0.3$ & $-0.3 \pm 0.6$   & $1.5\pm 0.2$ &  - & -   \\
 \hline
  Constant bias  & SDSS        & $-1.2\pm 0.4$  & $0.3\pm 0.6$   & $0.9\pm 0.2$ &  - & -   \\
    \hline
  Constant bias  & LRG + SDSS  & $-1.1 \pm 0.1$  & $0.5\pm 0.3$   & prior in bias          &  - & -   \\
   \hline
  Constant bias  & LRG + SDSS  & $-1.1 \pm 0.4$  & $0.3\pm 0.6$   & -          &  - & -   \\
   \hline
   \hline
 Evolved bias    &  LRG        & $-0.9\pm 0.3$ & $-0.2 \pm 0.5$    &  -         &  $1.7\pm 0.5$  & $(9.0 \pm 1.1)\times 10^{12}$   \\
 \hline
 Evolved bias    &  SDSS       & $-1.0 \pm 0.2$ & $0.6\pm0.4$   & -         &  $2.1\pm 0.6$  & $(9.0\pm1.5)\times 10^{11}$   \\
     \hline
 Evolved bias  & LRG + SDSS  & $-0.9 \pm 0.1$  & $0.3\pm 0.3$   &  -          &  prior in bias & prior in bias   \\
    \hline
 Evolved bias   & LRG+SDSS    & $-0.9 \pm 0.3$ & $0.2 \pm 0.5 $   & -         &  - & -   \\
  \hline
\end{tabular}
\caption{Best fit parameters of Dark Energy and Bias Model for ISW-g and galaxy-galaxy autocorrelation observations. The first column indicate the bias model we used. The second column shows the catalogs we used for the analysis. The third and forth columns are dark energy equation of state parameter $w_0$ and $w_a$. The fifth column is the constant bias parameter in constant bias model. The sixth and seventh column are the formation redshift and the mass of dark matter halo which hosts the galaxy, respectively. All the rows show the marginalized results for best fit parameters. Third and seventh rows are best fit parameters of DE with bias model parameters fixed. The LRG bias parameters are $b_{(lrg)}=1.8$, $z_{f(lrg)} = 2.0$ and $M_{(lrg)}= 10^{13}M_{\odot}$. The SDSS-galaxy free parameters are $b_{(sdss)}=1.0$, $z_{f(sdss)}=2.5$ and $M_{sdss}=10^{12} M_{\odot}$.}
\end{table}
As a final word, in this section we showed that the ISW-galaxy signal depends both on the cosmological model and the bias parameter which encapsulates in it the relation of luminous matter with underlying dark matter. Thus any deviation from $\Lambda$CDM prediction can be described with considering the physics of both phenomena. Therefore, theoretically motivated bias models like the evolution bias, which may be constrained with independent observations can break the degeneracy and help us to understand the cosmological models better.

\section{Conclusion and Future remarks}
\label{Sec:Conc}
The accelerated expansion of the universe is one of the greatest mysteries of modern cosmology. One way to address this problem is to measure the effect of the possible deviations from $\Lambda$, on observational data. The phenomenological parametrization of equation of state of cosmic fluid with a time dependent equation of state $w=w(z)$ is a way to address this problem via comparison with the cosmological data driven from geometrical and dynamical observations. One of the most prominent cosmological observations is ISW-g correlation which probes the geometry and dynamics of the cosmological models in different redshifts and in an integrated way. In this work, we show that the deviation of the cross correlation of the ISW-g signal from standard case is not proportional to deviation of the model in Poisson equation, (parameterized in $Q$-parameter), because it depends on the dynamical parameter of the ISW-g correlation and the distribution of the galaxies. Accordingly, we conclude that different galaxy catalogs with different selection functions can probe DE models in different redshifts. We show how the signal of ISW-galaxy correlation function depends on the joint kernel function ($W_{TM}$) which is obtained from dynamical parameters. The dynamical parameters are the growth function, the ISW kernel $W_T$, and the distribution of galaxies $dN/dz$.  \\
The other  important point we conclude is the degeneracy of the two different physics on the ISW-galaxy cross correlation. On one hand, we have the deviation from $\Lambda$CDM which affects the kernels and matter power spectrum respectively which can change the signal of the ISW-g correlation function. On the other is the bias parameter, the relation of luminous matter with the underlying distribution of dark matter. In this work we just assume the dark matter halo bias and we showed that the there is a degeneracy between two physics. \\
In this direction we studied the ISW-g signal with constant bias, the Press-Schechter bias and the evolved bias. We argue that the more accurate bias is the evolved one as the observation and formation time of dark matter halos do not coincide. It is shown that the evolved bias parameter is a probable solution to resolve the reported $1\sigma$ up to $2\sigma$ tension of the $\Lambda$CDM model with observational signals. This result is in agreement with recent work by Ferraro et al. \cite{Ferraro:2014msa} that assumes evolved bias which is fixed with lensing data. We also use only the ISW-galaxy data from SDSS galaxy sample and LRG to constrain the deviation from six-parameter $\Lambda$CDM model with the extension in equation of state and constant bias using phenomenological parametrization. We use the galaxy-galaxy auto-correlation function as a complementary observation to constrain the free parameters of DE and bias, simultaneously. In this direction, we used the constant bias model and the evolved bias model with two free parameters of halo mass and formation redshift. Marginalizing on DE parameters, we showed that the mass of LRG galaxies and formation redshift $M=10^{13} M_{\odot}$ , $z_{f}=2.0$ which is chosen in our case study analysis are consistent with data-fitting analysis in $2\sigma$ level. This is also true for SDSS galaxy sample with  $M=10^{12} M_{\odot}$ , $z_{f}=2.5$ .
We conclude that using the evolved bias tightens the constraint on the equation of state of  DE in comparison with the constant bias, in the case we can fix the bias from other observations. Otherwise, when we set the bias model parameters free, the above statement is not right.
The forecast of the constrains on the equation of state of DE could be an obvious extension of this work. In that case, if we assume, that the evolved bias is a more sophisticated model for bias which has its  free parameter as well, then we should find a way to break the degeneracy.
This is because the future galaxy surveys will extend the depth and width of the galaxy maps and the constraints on DE will be improved accordingly. Finally, two important points: a) The independent measurement of the bias parameter will have a crucial importance to pin down the dark matter baryonic segment of the problem and to improve the constraints on DE, b) Due to low signal to noise of ISW-LSS observation it will be crucial to add other cosmological observations like (CMB data, BAO, SNe, Weak Lensing) to this analysis to study the properties of DE.

\acknowledgments
We would like to thank Farshid Deylami for many discussions and comments on the work.
We would like to thank Tommaso Giannantonio for providing us with observational data of galaxy-galaxy auto-correlation.
Also we thank Yashar Akrami, Nima Khosravi and Sohrab Rahvar for insightful comments. We also thank the anonymous referee
which his/her detailed and insightful comments and suggestions helped us to improve the manuscript.


\end{document}